\documentclass[12pt, draftclsnofoot, journal, letterpaper, onecolumn]{IEEEtran}

\makeatletter
\def\ps@headings{%
\def\@oddhead{\mbox{}\scriptsize\rightmark \hfil \thepage}%
\def\@evenhead{\scriptsize\thepage \hfil \leftmark\mbox{}}%
\def\@oddfoot{}%
\def\@evenfoot{}}
\makeatother 

\IEEEoverridecommandlockouts

\usepackage{amsfonts}
\usepackage[dvips]{graphicx}
\usepackage{caption}
\usepackage{subfigure}
\usepackage{times}
\usepackage{cite}
\usepackage{lettrine}
\usepackage{amsmath}
\usepackage{amsmath}
\usepackage{amsthm}
\allowdisplaybreaks[4]
\usepackage{array}
\usepackage{amssymb}

\usepackage{stfloats}
\usepackage{slashbox}
\usepackage{graphicx}
\usepackage{footnote}
\usepackage{booktabs}
\usepackage{array}
\usepackage{algorithmic}
\usepackage{algorithm}
\usepackage{subeqnarray}
\usepackage{cases}
\usepackage{threeparttable}
\usepackage{color}

\newcommand{\bm}[1]{\mbox{\boldmath{$#1$}}}

\UseRawInputEncoding

\begin{document}

\title{Pricing-Driven Service Caching and Task Offloading in Mobile Edge Computing}

\IEEEoverridecommandlockouts

\author{Jia~Yan,~\IEEEmembership{Student Member,~IEEE}, Suzhi~Bi,~\IEEEmembership{Senior Member,~IEEE}, \\Lingjie Duan,~\IEEEmembership{Senior Member,~IEEE}, and Ying-Jun~Angela~Zhang,~\IEEEmembership{Fellow,~IEEE}
\thanks{Part of the work has been submitted to the IEEE International Conference on Communications (ICC), Montreal, Canada, June 14-18, 2021 \cite{ICC2021}.  J. Yan (yj117@ie.cuhk.edu.hk) and Y. J. Zhang (yjzhang@ie.cuhk.edu.hk) are with the Department of Information Engineering, The Chinese University of Hong Kong, Hong Kong.
S. Bi (bsz@szu.edu.cn) is with the College of Electronic and Information Engineering, Shenzhen University, Shenzhen, China.
L. Duan (lingjie$\_$duan@sutd.edu.sg) is with the Engineering Systems and Design Pillar, Singapore University of Technology and Design, Singapore. }
}
%\author{Jia ~Yan$^\dagger$, Suzhi~Bi$^*$, and Ying-Jun~Angela~Zhang$^\dagger$\\
%$^\dagger$Department of Information Engineering, The Chinese University of Hong Kong, Shatin, N.T., Hong Kong SAR\\
%$^*$College of Information Engineering, Shenzhen University, Shenzhen, Guangdong, China 518060\\
%E-mail:~ \{yj117, yjzhang\}@ie.cuhk.edu.hk, ~bsz@szu.edu.cn \vspace{-2ex}}

\maketitle

\vspace{-1.5cm}

\begin{abstract}
   Provided with mobile edge computing (MEC) services, wireless devices (WDs) no longer have to experience long latency in running their desired programs locally, but can pay to offload computation tasks to the edge server. Given its limited storage space, it is important for the edge server at the base station (BS)  to determine which service programs to cache by meeting and guiding WDs' offloading decisions. In this paper, we propose an MEC service pricing scheme to coordinate with the service caching decisions and control WDs' task offloading behavior in a cellular network. We propose a two-stage dynamic game of incomplete information to model and analyze the two-stage interaction between the BS and multiple associated WDs. Specifically, in Stage I, the BS determines the MEC service caching and announces the service program prices to the WDs, with the objective to maximize its expected profit under both storage and computation resource constraints. In Stage II, given the prices of different service programs, each WD selfishly decides its offloading decision to minimize individual service delay and cost, without knowing the other WDs' desired program types or local execution delays.  Despite the lack of WD's information and the coupling of all the WDs' offloading decisions, we derive the optimal threshold-based offloading policy that can be easily adopted by the WDs in Stage II  at the Bayesian equilibrium.  In particular, a WD is more likely to offload when there are fewer WDs competing for the edge server's computation resource, or when it perceives a good channel condition or low MEC service price.  Then, by predicting the WDs' offloading equilibrium, we jointly optimize the BS' pricing and service caching in Stage I via a low-complexity algorithm.  In particular, we study both the uniform and differentiated pricing schemes. For differentiated pricing, we prove that the same price should be charged to the cached programs of the same workload.
\end{abstract}
\begin{keywords}
Mobile edge computing, service caching and pricing, computation offloading, dynamic game under incomplete information.
\end{keywords}

%In this paper, we propose a mobile edge computing (MEC) service pricing scheme to coordinate with service caching decisions at base station (BS) and control wireless devices' (WDs) task offloading behavior. We propose a two-stage dynamic game of incomplete information to model and analyze the interaction between BS and WDs. In Stage I, the BS determines the MEC service caching and announces the service program prices to WDs, with the objective to maximize its expected profit under both storage and computation resource constraints. In Stage II, given the prices of different service programs, each WD selfishly decides its offloading decision to minimize individual service cost, without knowing the other WDs' desired program types or local execution delays. Despite the lack of WD's information and the coupling of all the WDs' offloading decisions, we derive the optimal threshold-based offloading policy in Stage II at the Bayesian equilibrium. Then, by predicting the WDs' offloading equilibrium, we jointly optimize the BS' pricing and service caching in Stage I via a low-complexity algorithm. In particular, we study both the uniform and differentiated pricing schemes. For differentiated pricing, we prove that the same price should be charged to the cached programs of the same workload.

\section{Introduction}

%\subsection{Motivation and Key Contributions}

Varieties of modern mobile applications, such as face recognition, online gaming and augmented reality, have recently emerged into our daily life. Wireless devices (WDs) equipped with low-performance computation units often experience long latency to  run these emerging computation-heavy applications.
%However, the scarcity in computation capabilities makes such requirement impractical.
%with the explosive growth of computing demands and low-performance wireless devices (WDs) in cellular networks, the scarcity in computation capabilities is becoming an impediment for designing efficient IoT systems.
Alternatively, mobile edge computing (MEC) is a promising solution to provide high-performance computing for the WDs \cite{MECsurvey1,MECsurvey2}.
%Specifically, with MEC, the WDs opportunistically offload tasks to their nearby edge servers according to the current system operating states, e.g., channel condition and available computing resource. Compared to traditional mobile cloud computing, MEC efficiently reduces the high overhead and long backhaul latency.
Instead of forwarding tasks to the remote data center as traditional mobile cloud computing does, the WDs are able to offload their tasks to nearby edge servers, which efficiently reduces the high overhead and long backhaul latency. The global edge computing market is expected to reach $\$28.07$ billion by 2027.\footnote{https://meticulousblog.org/top-10-companies-in-edge-computing-market/} For example, Amazon offers many MEC services, such as AWS IoT Greengrass\footnote{https://aws.amazon.com/greengrass/}, where users are charged based on the individual services they need.

%In order to improve the computing performance (i.e., minimum computation delay and energy consumption) in MEC,
In cellular networks, WDs can opportunistically offload their tasks to the edge server according to time-varying channel conditions and dynamic resource availability at the edge server. Most of the existing work on opportunistic computation offloading \cite{MEC2,MEC3,myfirstTWC,MEC1,partial2} assumes that the edge server has stored all the service programs. In practice, however, the service program acquiring process is time-consuming even during off-peak traffic hours. Compared with the task execution time at a millisecond level, the installation and loading time of a program takes tens of seconds for some common applications \cite{programinstall}. As such, the low latency requirement does not allow the edge server to fetch remotely from the program provider every time an MEC service is required. The edge server needs to pre-cache popular programs before repeatedly requested by WDs' offloaded tasks. Due to the limited caching storage capacity, the edge server needs to be selective and can only cache a subset of requested service programs before serving the WDs.
For the uncached service programs, the edge server is unable to provide the real-time computation services to the corresponding WDs' online applications.
This is referred to as \emph{service caching} \cite{service1,service2,service3,service4,service5,service6}.

In this paper, we consider an MEC system with a base station (BS) and multiple associated WDs. The MEC server is co-located with the BS. Specific service programs are required to compute the tasks for the WDs. For instance, a human face recognition service program at the edge server can be repetitively called to process individual pictures of different WDs. Based on its storage and computational capacity, the BS determines which service programs to cache and what prices to charge for the MEC service provided to the WDs. Based on the BS' service caching and pricing decisions as well as the competition from the other WDs, each WD decides whether to compute its task locally or at the edge server.

Making the optimal service caching, pricing, and offloading decisions is a challenging task. Intuitively, the BS needs to know the WDs' offloading decisions, so that it can cache the service programs that are most popularly requested. Likewise, each WD's offloading decision is affected by not only the service caching and pricing of the BS, but also the offloading decisions of the other WDs due to the sharing of MEC server resources. However, in practice the WDs are unwilling to reveal their private information about their local computing capabilities, task offloading delays and requested service program types. As such, their offloading decisions cannot be accurately inferred by the BS and the other WDs.

To address the above problem, we model the practical interaction between the BS and WDs as a two-stage dynamic game of incomplete information (a.k.a Stackelberg game under incomplete information) \cite{gamebook,pricingsurvey3}, where only the random distributions of each WD's characteristics, including requested service program type, local execution delay and offloading time, are known to the others.
%Note that pricing has been used as a promising way to balance the demand and supply in the capacity-limited service market  \cite{pricingsurvey1,pricingsurvey2}. We employ pricing to help manage the service caching decisions at the BS and guide task offloading from the WDs.
Specifically, in Stage I, the BS determines the service caching decisions and sets service prices for tasks requiring different programs to maximize its expected profit subject to the caching space and computation constraints. A higher price for a certain service program decreases the offloading willingness of the tasks requiring the program, but spares more computation resource to serve other types of tasks.
 %On the other hand, the BS can encourage offloading of a certain type of tasks by decreasing the price of the corresponding program or increasing the other programs' prices.
The prices for different service programs are highly related to the total caching space and computing power at the BS, the size of each service program, the popularity of each service program, and the WDs' willingness to offload their tasks under a certain price (which is not known precisely due to the lack of complete information). In this regard, we are interested in answering the first key question: \emph{What is the BS' optimal pricing and service caching strategy that maximizes its expected profit under incomplete information about the WDs?}

In Stage II, based on the given prices, the WDs compete for the computation resource at the BS and make task offloading decisions individually to minimize their own costs. Note that the WDs' optimal offloading decisions are coupled due to the sharing of the limited  resources at the edge server. Under such negative externalities, a WD may choose not to offload its task to the edge server if it predicts that many other WDs are going to offload, leaving the edge server little computing power to execute its task.  Accordingly, the second key question is raised: \emph{How should each WD decide its offloading decision and how different WDs (with the same or different programs) affect each other's decision-making under incomplete information?}

The main contributions in this paper are concluded as follows:
\begin{itemize}
  \item \emph{Two-stage dynamic game (a.k.a Stackelberg game) of incomplete information for managing service caching and task offloading:} To our best knowledge, this is the first work that studies two-stage dynamic game of incomplete information to jointly coordinate edge service caching and guide computation task offloading in the MEC systems. Besides selectively caching programs to admit the target WDs, we employ pricing \cite{pricingsurvey1,pricingsurvey2} as another degree of control to mitigate WDs' competition for limited computation resource and maximize the BS' profit. The BS can encourage offloading of a certain type of tasks by decreasing the price of the corresponding program or increasing the other programs' prices.
  \item \emph{Bayesian equilibrium of WDs' offloading decisions:}  For any given prices in Stage I, we analyze the WDs' optimal offloading decisions by considering their mutual competition and incomplete information. We show that each WD will follow a threshold-based task offloading policy at the Bayesian equilibrium, which is simple to implement in practice. More specifically, the threshold is a function of the programs' prices, the BS' CPU computation frequency, and the statistic characteristics of WDs' private information.
  \item \emph{Optimal strategy of BS' pricing and service caching:} Based on the analysis of the Bayesian equilibrium in Stage II, we derive two pricing schemes, namely, uniform pricing and differentiated pricing for Stage I. For uniform pricing, we suppose that the BS charges all the service programs with the same price and propose a low-complexity algorithm to jointly optimize the price and service caching decisions.
%We also shed light on the impact of various system parameters (e.g., edge computation capability) on the optimal service caching and task offloading solutions.
%We then study the optimal program pricing and service caching problem at the BS side. We first consider that the BS charges all the programs with the same price and derive the semi-closed-form solution for the optimal price. Accordingly, a low-complexity algorithm is proposed to optimize the common price and service caching decisions. We also shed light on the impact of various system parameters (e.g., edge computation capability) on the optimal service caching and task offloading solutions.
  Likewise, differentiated pricing assumes that the BS sets different prices for the service programs. Based on the analysis of the optimal prices for the cached service programs, an efficient optimization algorithm is proposed to obtain the optimal prices and service caching decisions. Interestingly, we show that the prices of two service programs are equal when they require the same computational workload. In a special case where the valuation of each WD's personalized information is uniformly distributed, we obtain more engineering insights on the optimal pricing.
\end{itemize}

%Simulation results verify our derived properties of the optimal solutions and show that our proposed pricing-based mechanism can efficiently control service caching and computation offloading in MEC.

The rest of the paper is organized as follows. In Section II, we introduce the system model. Section III formulates the two-stage dynamic game of incomplete information. We analyze the Bayesian subgame among the WDs in Stage II in Section IV. The uniform pricing scheme in Stage I is studied in Section V. We investigate the more general differentiated pricing scheme in Section VI. In Section VII, numerical results are described. Finally, we conclude the paper in Section VIII.

\subsection{Related Work}

Existing work has extensively studied opportunistic computation offloading, which is often jointly optimized with system computation and communication resource allocation \cite{MEC2,MEC3,myfirstTWC,MEC1,partial2}. Only recently has service caching started to attract research interests \cite{service1,service2,service3,service4,service5,service6}. For a single-server MEC system, \cite{service1} proposed an online algorithm to dynamically schedule the cached services without the knowledge of task arrival patterns.
%When the computation requires an uncached service, the edge server can either forward request to the cloud or download and install the whole service from the cloud at the cost of significant processing delay.
For a multi-server MEC system, \cite{service2} studied the joint service caching and request scheduling problem.
%, where edge servers cooperatively cache the requested service programs to serve the WDs.
The problem of minimizing served traffic load was considered in \cite{service3}.
\cite{service4} posed the service caching problem as a combinatorial bandit learning problem.
%, where a service provider dynamically rents computing and storage resources of the edge nodes to meet users' requests.
In \cite{service5}, each WD can offload its task to either the remote cloud center or a nearby edge node that has cached the required service program. Based on this, a joint optimization of service caching and task offloading was studied therein. Notice that the above work \cite{service1,service2,service3,service4,service5} has assumed that all tasks are computed at the edge server or/and remote cloud, neglecting the benefits of opportunistic computation offloading in MEC. For example, WDs may choose to compute locally when they perceive poor channel conditions. Very recently, \cite{service6} considered the joint optimization of service caching placement, task offloading, and resource allocation in a sequential task graph.

The above work has optimized the service caching, task offloading and resource allocation in a centralized manner. In terms of decentralized operation, previous work has proposed  Stackelberg games \cite{price1,price4}, priority pricing \cite{price2}, multi-round resource trading \cite{price3} in MEC systems. Specifically, \cite{price1} considered a Stackelberg game, where the edge server acts as the leader and sets prices to maximize its revenue with computation capacity constraint. The WDs are the followers and locally make offloading decisions to minimize their own costs for given prices. \cite{price4} studied optimal pricing and edge node selection by adopting a Stackelberg game. \cite{price2} further proposed a priority pricing scheme, where users are served first for a higher price. The authors in \cite{price3} designed an online multi-round auction mechanism for profit maximization.

\cite{price1,price4,price2,price3} have assumed that all programs are cached at the BS and  none of these studies have taken service caching into account. In this paper, we endeavor to design a Stackelberg game in MEC to coordinate the service caching and pricing decisions at the BS and the offloading behavior of the WDs. Besides, the existing work (e.g., \cite{price1,price4,price3}) has assumed that each WD's private information is known to all. In contrast, we take into account the information uncertainty when designing and analyzing the two-stage Stackelberg game.

\section{System Model}

\begin{figure}[t]
\begin{centering}
\includegraphics[scale=0.6]{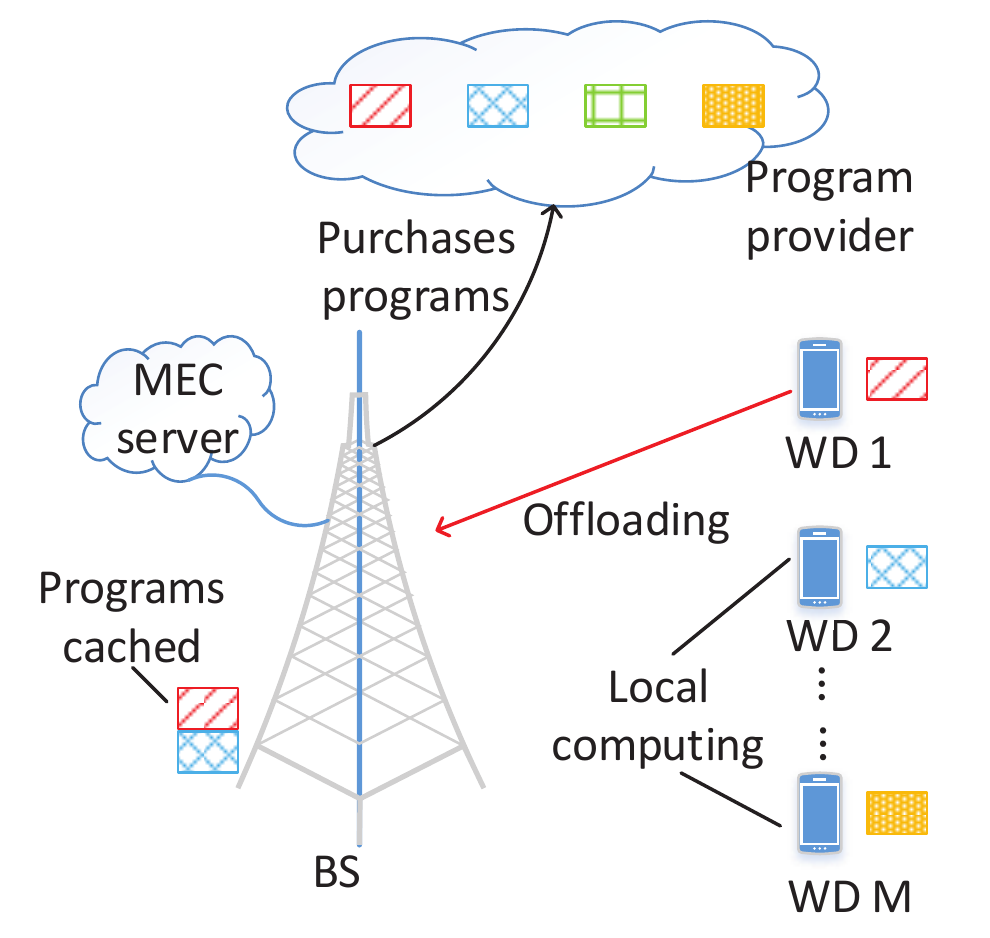}
\vspace{-0.1cm}
 \caption{System model of BS provision of MEC services to $M$ WDs who can choose to offload their desired programs' tasks to the edge server (i.e., BS) or compute locally. Given its limited caching storage, the BS can only serve those WDs if it caches such desired programs from the program provider beforehand. The BS cannot cache all potential programs from the program provider.}\label{fig1}
\end{centering}
\vspace{-0.1cm}
\end{figure}

As shown in Fig. \ref{fig1}, we consider a multi-user MEC system with $M$ single-antenna active WDs, denoted by a set $\mathcal{M}=\{1,2,...,M\}$, and one single-antenna BS. MEC server is located at the BS to share the infrastructure such as stable power supply. The BS provides MEC services to the WDs with its limited computation resource and storage capacity.
%, where the WDs can choose local execution or task offloading under desired programs to the BS. If the BS pre-caches those desired programs from the program provider, the corresponding WDs can be served.

%Each WD running one program (e.g., human face recognition) locally  has a computationally intensive task to execute now, and there are in total $N$ service programs potentially provided to the WDs by the program provider in Fig. 1, indexed by $\mathcal{N}=\{1,2,...,N\}$.

Suppose that each WD has a computationally intensive task to compute. The computation of each task requires a service program, e.g., human face recognition program. We refer to a task as a type-$j$ task if it is processed by the service program $j, j\in \mathcal{N}=\{1,2,...,N\}$, where $N$ is the total number of service programs. We define a binary indicator $u_{i,j}$ such that $u_{i,j}=1$ when the task of WD $i$ is of type $j$, and 0 otherwise. Accordingly, $\sum_{j}u_{i,j}=1, \forall i\in\mathcal{M}$, implying that a WD cannot run two programs at the same time. Besides, we denote the type of the WD $i$'s task as $\varphi_i\in\{1,...,N\}$. In particular, $\varphi_i=j$ if $u_{i,j}=1$.

Each WD needs to decide whether to compute its task locally or remotely at the BS. Define a binary indicator variable $a_i$ such that $a_i=1$ when WD $i$ decides to offload its task for edge computing and $a_i=0$ when WD $i$ decides to compute the task locally. A task can be served at the BS only if the BS has pre-cached the corresponding service program from the program provider. Take Fig. 1 for example. The BS has cached the required programs of WD 1 and WD 2. After comparing the costs of local computing and edge computing, WD 1 decides to offload its task for edge computing, while WD 2 decides to compute locally despite the availability of its required service program at the BS. WD $M$, on the other hand, has no choice but to compute locally, because its required service program is not cached at the BS. To avoid trivial cases, we assume that the WDs have the service programs to run their own tasks locally. Otherwise, they will always offload the tasks to the edge server.
%For brevity, we use the terms ``service program" and ``program" interchangeably in this paper.
In the following, we introduce the service caching, communication, and computation models in detail.

\subsection{BS' Service Caching Model}

Suppose that the cost for the BS to acquire the $j$-th service program from the program provider is $r_j$. After obtaining the program data and configuration from the program provider, the edge server installs and caches the service programs (e.g., executable .EXE files). We use a binary indicator $x_j$ to denote the caching decision of the $j$-th program at the BS. Specifically, $x_j=1$ tells that the $j$-th program is cached at the BS, and $x_j=0$ otherwise.
%Notice that the BS can execute a type-$j$ task only if the $j$-th program is cached.
Given the limited storage space, the BS cannot cache all the potential programs. We model the caching capacity constraint at the BS as
\begin{align}\label{cache_limit}
\sum_{j=1}^{N}x_jc_j\leq C,
\end{align}
where $c_j$ is the size of the $j$-th generated program and $C$ is the caching space at the BS. Besides, the BS needs the input task data (e.g. individual photos for the human face recognition program) from the WDs to run the cached programs.
%As shown in Fig. 1, if $x_j=1$, the BS pays the program provider $r_j$ to obtain the data of $j$-th program, then generates and caches the corresponding program. The BS can receive the computation offloading of type-$j$ tasks only if the program is cached.

%When the $i$-th WD chooses to offload its type-$j$ task to the edge server, the associated $j$-th program needs to be cached in the BS. That is, for the WD $i$, if $a_i=1$ and the task of WD $i$ is type-$j$, i.e., $u_{i,j}=1$, the BS should purchase the $j$-th program data with the payment $r_j$ from the program provider and cache the generated $j$-th program, i.e., $x_j=1$. Equivalently, the above relation between the offloading decision and caching decision is expressed as
%%
%\begin{align}
%a_i\leq\sum_{j=1}^{N}u_{i,j}x_j, \forall i\in\mathcal{M}.
%\end{align}
%%

\subsection{WDs' Communication Model with BS}

We assume that the WDs access the uplink spectrum through FDM or
OFDM to avoid mutual interferences. Each WD is fairly allocated an orthogonal channel of identical bandwidth $W$.\footnote{Note that there are mature ways such as spectrum management to control interference, which is out of the scope of this paper.} Let $p_i$ denote the transmit power of WD $i$ when offloading its task to the BS. The wireless channel gain between WD $i$ and the BS is denoted as $h_i$.
% and we assume offloading/downloading channel reciprocity in this paper.
Besides, we assume additive white Gaussian noise (AWGN) with zero mean and identical variance $\sigma^2$ at all the receivers. The offloading data rate of the task from WD $i$ to the BS is
\begin{align}
R_{i}^{u}=W\log_{2}\left(1+\frac{p_ih_i}{\sigma^2}\right).
\end{align}
Then, the transmission time of WD $i$ when offloading its task is expressed as
\begin{align}\label{offloading}
\tau_{i}^{u}=\frac{I_i}{R_{i}^{u}},
\end{align}
where $I_i$ is the size (in bits) of the input data of WD $i$'s task. In this paper, we suppose that different computation tasks under the same service program can have different data inputs and outputs. For example, when running the human face recognition application, WDs need to input their photos of different sizes and definitions and expect the program to return specific results.
%
%and the corresponding energy consumption of the WD $i$ is
%%
%\begin{align}
%e_{i}^{u}=p_i\tau_{i}^{u}.
%\end{align}
%%

Finally, we assume that the time spent on downloading the task computation result from the BS to the WD is negligible due to the strong transmit power of the BS and the relatively small output data size (as compared to the input data size). For instance, the human face recognition application outputs the person name with only a few bytes, which is much smaller than the size of the corresponding input photo (of several mega bytes).

\subsection{Computation Model at the BS and the WDs}

If WD $i$ computes its task locally, i.e., $a_i=0$, then the local execution time is
\begin{align}\label{local}
\tau_{i}^{l}=\frac{\sum_{j=1}^{N}u_{i,j}L_j}{f_i^l},
\end{align}
where $f_i^l$ is the CPU computation frequency of WD $i$ and $L_j$ denotes the computational workload in CPU cycles to execute the type-$j$ task.
%and the corresponding energy consumption is given by
%%
%\begin{align}
%e_{i}^{l}=\kappa L_i(f_{i}^l)^{2}=\kappa\frac{L_i^3}{(\tau_i^l)^2},
%\end{align}
%%
%where $\kappa$ is the computing energy efficiency parameter depending on the chip architecture.

Alternatively, WD $i$ can choose to offload its task in MEC services, i.e., $a_i=1$. Suppose that the edge server creates multiple virtual machines (VMs) to execute the offloaded tasks in parallel and each VM is assigned to handle one task.  For simplicity, we assume that the total computation resource at the edge server (i.e., the total CPU frequency $f^c$) is equally partitioned and allocated to the VMs. Accordingly, the task processing time at the edge server is
\begin{align}\label{edge}
\tau_{i}^{c}(m)=\frac{\sum_{j=1}^{N}u_{i,j}L_j}{f^c/m},
\end{align}
where $m$ is the number of WDs' tasks offloaded to the edge server for edge computing, i.e., $m=\sum_{i=1}^Ma_i$. Notice that $m$ is a random variable depending on WDs' offloading decisions $a_i$.  $\tau_{i}^{c}(m)$ is an increasing function of $m$.
%the service rate for task of WD $i$ at the edge is modeled as
%%
%\begin{align}
%\tilde{f}_{i}^{c}=\frac{f^c}{m},
%\end{align}
%%

%We consider the I/O interference for parallel computing in the edge server, where the service rate $\tilde{f}_i^c$ of the VM executing task $i$ suffers from a degradation when multiplexed with another VMs. Suppose that the expected CPU frequency of the VM executing task $i$ when running in isolation is $f_i^c$. Without loss of generality, we assume that $f_i^c=f^c, \forall i,$ for brevity.  Then, the degraded service rate for task $i$ at the edge is modeled as
%
%\begin{align}
%\tilde{f}_{i}^{c}=f^c(1+\gamma)^{1-m},
%\end{align}
%
%where $\gamma>0$ is the degradation factor corresponding to the I/O interference at the edge server, and $m$ is the number of tasks offloaded to the edge for parallel computing.

%Let $D$ denote the energy consumption per cycle for edge computing. Then the product $DL_i$ gives the computing energy for the task of WD $i$ at the edge side, i.e.,
%%
%\begin{align}
%e_{i}^{c}=DL_i.
%\end{align}
%%

\section{Two-Stage Dynamic Game Formulation under Incomplete Information for MEC Service Provision}

\begin{figure}[htb]
\begin{centering}
\includegraphics[scale=0.6]{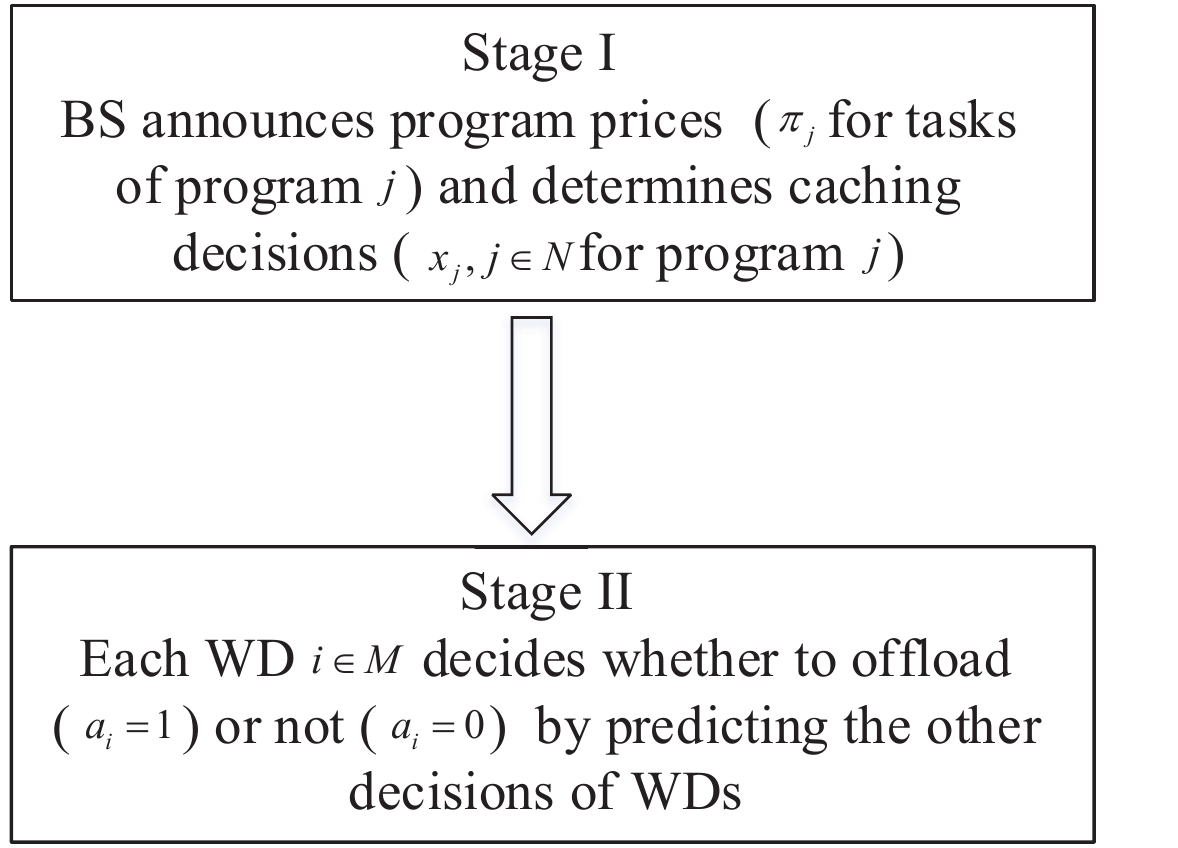}
\vspace{-0.1cm}
 \caption{Our proposed two-stage dynamic game for the interaction between the BS and the WDs. }\label{fig2}
\end{centering}
\vspace{-0.1cm}
\end{figure}

%In our considered MEC system model, the BS prices and provides services and the WDs' selfish offloading decisions depend on the service caching decision and communication/computation capability as well as the prices charged at the BS.
The BS and the WDs interact with each other in a two-stage dynamic game of incomplete information as shown in Fig. \ref{fig2}, where the BS is the leader and the WDs are the followers.
The profit-seeking BS first determines the caching decisions and sets the  program prices for task executions in Stage I. Then, the delay-sensitive WDs optimize their offloading decisions individually in Stage II based on the prices announced from the BS and the WDs' mutual competition to share the limited computation resource at the BS. We will analyze this dynamic game by backward induction. In the following, we first detail the problem formulation in each stage.

\subsection{WDs' Problem Formulation in Stage II}

Each WD aims to make an optimal offloading decision to minimize its total cost, defined as its task execution delay plus the payment to the BS.  In particular, the task execution time of WD $i$, denoted by $T_i$, is
\begin{align}\label{delay}
T_i(m)=(1-a_i)\tau_i^l+a_i(\tau_i^u+\tau_i^c(m)).
\end{align}
Note that if $a_i=0$, the total delay $T_i$ in \eqref{delay} simply equals the local execution time $\tau_i^l$ in \eqref{local}. Otherwise, $T_i$ consists of the data uploading time $\tau_i^u$ in \eqref{offloading} and the edge computing time $\tau^c_i$ in \eqref{edge}, which increases with $m$, the total number of tasks offloaded to the BS.
If WD $i$ chooses to offload its task, i.e., $a_i=1$, then it needs to pay the BS for the service. Suppose that the price the BS charges for program $j$'s task execution is $\pi_j$ per CPU cycle. Then, the total amount WD $i$ pays is $a_i\sum_{j=1}^Nu_{i,j}L_j\pi_j$. As such, the total cost WD $i$ aims to minimize is
\begin{align}
U_i(m)=T_i(m)+a_i\sum_{j=1}^Nu_{i,j}L_j\pi_j.
\end{align}

In an ideal case where WD $i$ has the knowledge of $m$, it can simply compare $U_i(m|a_i=1)$ and $U_i(m|a_i=0)$ and choose the offloading decision that yields the smaller value of $U_i$.
In practice, however, a WD is not able to infer the other WDs' offloading decisions, as their private information including local computing time $\tau_i^l$, offloading time $\tau_i^u$, computation workload $L_{\varphi_i}$ and program type $u_{i,j}$ is not revealed. As such, the exact value of $m$ is unknown. To address the issue, we will derive the Bayesian equilibrium to understand each WD's best response under incomplete information in Section IV.

%However, this decision-making requires knowledge of $m$ and is not easy to decide under incomplete information about the other WDs' decisions and personal information. Here, we practically consider that all the private information including local computing time $\tau_i^l$, offloading time $\tau_i^u$, computation workload $L_j$ and program type $u_{i,j}$ are only known by WD $i$ and is unknown by the other WDs and the BS. In this case, each WD does not know the other WDs' offloading decisions and thus the exact number $m$ of tasks offloaded to the edge.

\subsection{BS' Problem Formulation in Stage I}
The BS aims to maximize its overall profit by optimizing the service caching $\mathbf{x}=\{x_j,j\in\mathcal{N}\}$ and program pricing $\bm{\pi}=\{\pi_j,j\in\mathcal{N}\}$. In particular, the profit is the difference between the payments received from the WDs and the cost of acquiring the programs from the program provider:
%At the BS side, the target is to determine the service caching decisions and set the program prices for task executions to guide WDs' competitive offloading behavior and thus maximize the BS' total revenue.  The BS charges WD $i$ the amount $a_i\sum_{j=1}^Nu_{i,j}L_j\pi_j$ only after the corresponding service program is cached. Accordingly, the revenue of the BS is given by
%
\begin{align}\label{uB}
u_B=\sum_{i=1}^Ma_i\sum_{j=1}^Nx_ju_{i,j}L_j\pi_j-\sum_{j=1}^Nx_jr_j.
\end{align}
The first term in the right hand side of \eqref{uB} is the total payments received from the WDs. Here, the multiplicative factor $x_j$ corresponds to the fact that the BS can serve a type-$j$ task only when the $j$-th program is cached. The second term in \eqref{uB} is the total program acquiring cost from the program provider.

Under the assumption of incomplete information, the BS does not know each WD's private information, including local execution time $\tau_i^l$, offloading time $\tau_i^u$, computation workload $L_{\varphi_i}$ and program type $u_{i,j}$. Instead, the BS only knows the distribution of each WD's characteristics. As such, the BS infers the offloading probabilities for the WDs and computes the expected profit as
\begin{align}\label{UB}
U_B=\mathbb{E}\left[u_B\right]=\mathbb{E}\left[\sum_{i=1}^Ma_i\sum_{j=1}^Nx_ju_{i,j}L_j\pi_j-\sum_{j=1}^Nx_jr_j\right],
%&=\sum_{j=1}^N[1-F(\delta+\pi_j)]x_jq_{j}ML_j\pi_j-\sum_{j=1}^Nx_jr_j,
\end{align}
where the expectation is taken over the distributions of service program types $u_{i,j}$ and the offloading decisions $a_i$ the WDs made in the Stage II game.

%The goal of the BS is to set the optimal prices $\bm{\pi}=\{\pi_j,j\in\mathcal{N}\}$ and caching decisions $\mathbf{x}=\{x_j,j\in\mathcal{N}\}$ to maximize its expected revenue given its caching capacity constraint.
Mathematically, the optimization problem at the BS is formulated as
\begin{eqnarray}
\mbox{(P1)}~~\max_{(\bm{\pi},\mathbf{x})}&&U_B,\nonumber\\
{\rm s.t.}&&\sum_{j=1}^{N}x_jc_j\leq C,\nonumber\\
&&x_{j}\in\{0,1\},\forall j=1,...,N.
\end{eqnarray}

In the following, we analyze the proposed two-stage dynamic game using backward induction. In Section IV, we first start with the Stage II game, when the service price $\bm{\pi}$ is fixed. In particular, we will analyze the Bayesian subgame among the WDs and obtain the equilibrium decision policy for all the WDs. Then in Section V and VI, we analyze the Stage I where the BS optimizes programs' prices $\bm{\pi}^*$ and caching decisions $\mathbf{x}^*$ to maximize its expected profit, by predicting the WDs' equilibrium offloading behavior.

\section{Analysis of the WDs' Offloading Equilibrium in Stage II}

By observing the prices announced by the BS in Stage I, the WDs determine the offloading decisions individually by estimating the other WDs' decisions, which leads to a Bayesian subgame in Stage II. An equilibrium is reached if no WD can improve its cost by changing its offloading strategy unilaterally. In this section, we manage to derive an optimal threshold-based offloading strategy for all the WDs at the Bayesian subgame equilibrium (as a stable outcome of the WDs' interactions) and analyze some interesting properties of the optimal threshold.

To derive the Bayesian subgame equilibrium under incomplete information, we first try to understand the properties of the optimal offloading decisions in an ideal case where the private information of each WD is known by each other. That is, each WD exactly knows the number $m$ of tasks offloaded to the BS by calculating the offloading decisions on other WDs' behalf.

\emph{\textbf{Lemma 4.1 (complete information scenario):}} The optimal offloading decision of WD $i$ with type-$\varphi_i$ task is given by
\begin{align}\label{lemma41}
a_i^*(\pi_{\varphi_i})=\left\{
      \begin{array}{ll}
        1, & \pi_{\varphi_i}\leq\theta_i-\frac{m}{f^c}; \\
        0, & \mbox{otherwise} ,
      \end{array}
    \right.
\end{align}
where
\begin{align}\label{definetheta}
\theta_i=\frac{\tau_i^l-\tau_i^u}{L_{\varphi_i}}.
\end{align}
%

%Note that the valuation $\theta_i$ represents WD $i$'s willingness-to-offload.
Lemma 4.1 implies that WD $i$ will choose to offload if its total payment is smaller than the difference between the local execution time and the edge computing time, i.e., $L_{\varphi_i}\pi_{\varphi_i}\leq L_{\varphi_i}\theta_i-\frac{mL_{\varphi_i}}{f^c}$. Note that the WD $i$'s willingness-to-offload increases when its local computation time $\tau_i^l$ is long, the data uploading to the BS incurs short delay (small $\tau_i^u$), or there are a smaller number $m$ of WDs competing for the BS' computation resource $f^c$.

Now we turn to the incomplete information scenario where neither $m$ nor $\theta_i$ is publicly known. Instead, $\theta_i$ appears random to other WDs and the BS. We assume that $\theta_i$'s are independent and identically distributed with probability density function (PDF) $f(\cdot)$ and cumulative distribution function (CDF) $F(\cdot)$. The distribution function is a common prior knowledge to all the WDs and the BS. As in the wide literatures of pricing and mechanism designs \cite{regular}, we suppose that the distribution of $\theta_i$ is regular as defined below.

\emph{\textbf{Assumption 1 (regular distribution):}} $y(\theta)=\theta-\frac{1-F(\theta)}{f(\theta)}$ is an increasing function of continuous random variable $\theta$, where $F(\theta)$ and $f(\theta)$ are the CDF and PDF of $\theta$, respectively.

Note that many random distributions, such as uniform, normal, and exponential distributions, are indeed regular distributions. Likewise, we assume that WD $i$ does not know the other WDs' task types, so that the prices that the other WDs need to pay for offloading are unknown. In this regard, we assume that the program indicator $u_{i,j}$ for each WD $i$ appears random to other WDs with probability $q_j$. In particular, $q_j$ represents the $j$-th program's popularity and can be estimated by all the WDs and BS by observing the historical data.

Then, the PDF of $\beta_i=\theta_i-\pi_{\varphi_i}$ is given by
\begin{align}
g(\beta_i)=\sum_{j=1}^{N}q_jf(\beta_i+\pi_j),
\end{align}
and the corresponding CDF $G(\beta_i)$ is
\begin{align}\label{CDF_G}
G(\beta_i)&=\int_{-\infty}^{\beta_i}g(x)dx=\int_{-\infty}^{\beta_i}\sum_{j=1}^{N}q_jf(x+\pi_j)dx=\sum_{j=1}^{N}q_jF(\beta_i+\pi_j).
\end{align}

Using the definition above and based on Lemma 4.1, we obtain the optimal offloading strategy for each WD at the Bayesian equilibrium under the incomplete information scenario in the following Theorem 1.

\emph{\textbf{Theorem 1 (incomplete information scenario):}} A WD $i$ with type-$\varphi_i$ task will offload its task to the BS if and only if
\begin{align}
\pi_{\varphi_i}\leq\theta_i-\delta^*(\bm{\pi}).
\end{align}
Here, the equilibrium decision parameter $\delta^*(\bm{\pi})$ is the same for all the WDs and is the unique solution to
\begin{align}\label{stage2}
\Phi(\delta):=\delta-\frac{(M-1)(1-G(\delta))+1}{f^c}=0.
\end{align}
Besides, $\delta^*(\bm{\pi})$ satisfies $\frac{1}{f^c}\leq\delta^*(\bm{\pi})\leq\frac{M}{f^c}$.
%where the expectation is taken over $m'$ which follows a binomial distribution $B(M-1,1-G(\delta))$.

\begin{proof}
Due to symmetry, we assume that all the WDs other than WD $i$ choose to offload their tasks to the edge server if and only if their valuations $\beta_{k}, k\neq i,$ are larger than a decision parameter $\delta>0$. By extending Lemma 4.1 to the incomplete information case, we know that when the following inequality holds, WD $i$ would prefer to offload its task for edge computing.
\begin{align}\nonumber
\theta_i-\pi_{\varphi_i}\geq\mathbb{E}\left[\frac{m'+1}{f^c}\right],
\end{align}
where $m'$ follows a binomial distribution $B(M-1,1-G(\delta))$ and represents the number of the WDs other than WD $i$ that prefer edge computing. Here, $\mathbb{E}\left[\frac{m'+1}{f^c}\right]=\frac{(M-1)(1-G(\delta))+1}{f^c}$. At the equilibrium, we have the common decision parameter $\delta$. That is,
\begin{align}\nonumber
\delta= \frac{\mathbb{E}_{m'}\left[m'\right]+1}{f^c}=\frac{(M-1)(1-G(\delta))+1}{f^c}.
\end{align}

Next, we want to show that there exists a unique solution $\delta^*$ to \eqref{stage2}. %We define
%%
%\begin{align}
%\Phi(\delta)=\delta-\frac{(M-1)(1-G(\delta))+1}{f^c}.
%%&=\sum_{m'=0}^{M-1}\left(\frac{(1+\gamma)^{m'}}{f^c}\right)\tbinom{M-1}{m'}\left(1-G(\delta)\right)^{m'}\left(G(\delta)\right)^{M-1-m'}.
%\end{align}
Since $G(\delta)$ in \eqref{CDF_G} is an increasing function with respect to $\delta$, $\Phi(\delta)$ is a monotonically increasing function with respect to $\delta$, i.e., $\frac{\partial G(\delta)}{\partial\delta}>0$. When $\delta=\frac{1}{f^c}$, we have $\Phi(\delta)\leq 0$. When $\delta=\frac{M}{f^c}$, we have $\Phi(\delta)\geq 0$. Together with the result that $\Phi(\delta)$ is a monotonically increasing function, $\Phi(\delta)=0$ has a unique solution $\delta^*\in[\frac{1}{f^c},\frac{M}{f^c}]$.
\end{proof}

According to Theorem 1, we can obtain $\delta^*$ through a bi-section search over $\delta^*\in [\frac{1}{f^c},\frac{M}{f^c}]$ that satisfies $\Phi(\delta^*)=0$. $\delta^*(\bm{\pi})$ can be viewed as the expectation of $\frac{m}{f^c}$ in \eqref{lemma41} under incomplete information. Despite WDs' heterogeneity in local execution time, communication delay and program type, the proposed policy integrates all such personal information in a single parameter $\theta_i$. To decide whether to offload, a WD just needs to compare its private term $\theta_i$ with the decision threshold, which is defined as the equilibrium parameter $\delta^*$ plus price $\pi_{\varphi_i}$ for its desired program (i.e., $\theta_i\geq\pi_{\varphi_i}+\delta^*(\bm{\pi})$).
%It is worth noting that the offloading decision of WD $i$ with type-$j$ task is jointly determined by the common equilibrium threshold $\delta$ and the price $\pi_j$. We denote the decision threshold for WD $i$ as
%%
%\begin{align}
%\delta_i=\delta+\Sigma_{j=1}^{N}u_{i,j}\pi_j.
%\end{align}
%%
%Accordingly, the WD $i$ will offload its task to the edge if the valuation $\theta_i$ is larger than $\delta_i$.
In the following, we derive some interesting properties of decision threshold $\delta^*+\pi_{\varphi_i}$ for WD $i$.

\emph{\textbf{Proposition 4.1:}} The WD $i$'s decision threshold $\delta^*+\pi_{\varphi_i}$ increases in its own program's price $\pi_{\varphi_i}$ and decreases in any other program's price $\pi_{k}, k\in\mathcal{N}\setminus \varphi_i$.

\begin{proof}
Please refer to Appendix \ref{appendicesA}.
\end{proof}

The Proposition 4.1 indicates that the BS can incentivize the WDs with type-$j$ tasks to offload by setting a lower price for the $j$-th program or a higher price for the other programs.
%It is because from the BS's viewpoint, decreasing $\delta_i$ leads to a larger offloading probability (i.e., $1-F(\delta_i)$) for the WD $i$ under the incomplete information scenario.
Next, we further study the impacts of the edge server's CPU computation frequency $f^c$ and the total number of users $M$ on the offloading decision threshold $\delta^*+\pi_{\varphi_i}$.

\emph{\textbf{Proposition 4.2:}} The WD $i$'s decision threshold $\delta^*+\pi_{\varphi_i}$ decreases in $f^c$, and increases in $M$.

\begin{proof}
Please refer to Appendix \ref{appendicesB}.
\end{proof}

It follows from Proposition 4.2 that WDs are more likely to offload when the BS has larger computational capability. Besides, as more WDs compete for the limited computation resource, each WD tends to offload with lower probability to avoid long computation latency at the BS.

By predicting the WDs' equilibrium strategies in Stage II through Theorem 1, we are ready to turn to Stage I in the following section.

\section{Uniform Pricing to Cope with MEC Service Caching}

The optimal decision in Stage I involves combinatorial optimization due to the binary service caching decisions $\mathbf{x}$ in (P1). In this section, we first propose an efficient algorithm to optimize the program prices and service caching decisions under a uniform pricing strategy, where the BS sets the same unit price to all the programs, i.e., $\pi=\pi_1=...=\pi_N$. The identical pricing among all the programs simplifies our analysis of Problem (P1).

\subsection{BS' Profit Maximization in Problem (P1)}

Based on the Bayesian equilibrium policy derived in Stage II, the BS can effectively predict the WDs' offloading behaviors by calculating the offloading probability of each program's tasks. Specifically, a task is offloaded with probability $1-F(\delta^*(\pi)+\pi)$ regardless of its type due to the uniform pricing. Then, we can rewrite Problem (P1) in Stage I as
\begin{eqnarray}
\mbox{(P2)}~~\max_{(\pi,\mathbf{x})}&&(1-F(\delta^*(\pi)+\pi))\pi\sum_{j=1}^N L_jx_jq_jM-\sum_{j=1}^Nx_jr_j,\nonumber\\
{\rm s.t.}&&\sum_{j=1}^{N}x_jc_j\leq C,\nonumber\\
&&x_{j}\in\{0,1\},\forall j=1,...,N.
\end{eqnarray}

Note that the optimization for the price $\pi$ is decoupled from the caching decisions $\mathbf{x}$ in Problem (P2). Thus, we can separately  optimize $\pi$ by maximizing the term $(1-F(\delta^*(\pi)+\pi))\pi$ in the objective function of Problem (P2). In particular, the following Proposition 5.1 shows that the optimal price $\pi^*$ can be efficiently obtained using bi-section search method.

\emph{\textbf{Proposition 5.1:}} The optimal uniform price $\pi^*$ is the unique solution to
\begin{align}\label{uniformopt}
\omega_u(\pi):=\pi-\frac{\left[1-F(\delta^*(\pi)+\pi)\right]}{f(\delta^*(\pi)+\pi)}-\frac{(M-1)\left[1-F(\delta^*(\pi)+\pi)\right]}{f^c}=0,
\end{align}
where $\delta^*(\pi)$ is given in \eqref{stage2}.

\begin{proof}
We first prove the optimality of the solution in \eqref{uniformopt}.
The derivative of $(1-F(\delta^*(\pi)+\pi))\pi$ with respect to $\pi$ is
\begin{align}\label{ubpipro6}
\frac{\partial [(1-F(\delta^*(\pi)+\pi))\pi]}{\partial\pi}=\left[\pi\left[-f(\delta^*+\pi)(\frac{\partial\delta^*}{\partial\pi}+1)\right]+\left[1-F(\delta^*+\pi)\right]\right].
\end{align}

According to the analysis in Stage II, we have
\begin{align}\label{deltapi}
\frac{\partial\delta^*}{\partial\pi}=\frac{-(M-1)f(\delta^*+\pi)}{f^c+(M-1)f(\delta^*+\pi)}.
\end{align}

By substituting \eqref{deltapi} into \eqref{ubpipro6} and letting $\frac{\partial U_B}{\partial\pi}=0$, we have
\begin{align}
\pi^*-\frac{\left[1-F(\delta^*+\pi^*)\right]}{f(\delta^*+\pi^*)}\frac{f^c+(M-1)f(\delta^*+\pi^*)}{f^c}=0,
%&=\frac{\left[1-F(\delta^*+\pi^*)\right]}{f(\delta^*+\pi^*)}+\frac{(M-1)\left[1-F(\delta^*+\pi^*)\right]}{f^c}.\nonumber
\end{align}
which yields $\omega_u(\pi^*)=0$.

Next, by analyzing the property of $\omega_u(\pi)$, we demonstrate the existence and uniqueness of the optimal price. According to Theorem 1, we rewrite $\omega_u(\pi)$ as
\begin{align}\label{wu}
\omega_u(\pi)=\delta^*(\pi)+\pi-\frac{\left[1-F(\delta^*(\pi)+\pi)\right]}{f(\delta^*(\pi)+\pi)}-\frac{2(M-1)\left[1-F(\delta^*(\pi)+\pi)\right]+1}{f^c}.
\end{align}
Based on Assumption 1, $\delta^*(\pi)+\pi-\frac{\left[1-F(\delta^*(\pi)+\pi)\right]}{f(\delta^*(\pi)+\pi)}$ in \eqref{wu} is an increasing function in $\delta^*(\pi)+\pi$. Besides, the last term $-\frac{2(M-1)\left[1-F(\delta^*(\pi)+\pi)\right]+1}{f^c}$ in \eqref{wu} is an increasing function in $\delta^*(\pi)+\pi$. Therefore, all the terms in $\omega_u(\pi)$ increase with $\delta^*(\pi)+\pi$, implying that $\omega_u(\pi)$ increases in $\delta^*(\pi)+\pi$. Then, according to Proposition 4.1, $\delta^*(\pi)+\pi$ increases in $\pi$. Thus, we have $\omega_u(\pi)$ is an increasing function in $\pi$. Meanwhile, when $\pi=0$, we have
\begin{align}
\omega_u(\pi=0)=-\frac{\left[1-F(\delta^*)\right]}{f(\delta^*)}-\frac{(M-1)\left[1-F(\delta^*)\right]}{f^c}<0.\nonumber
\end{align}
When $\pi\rightarrow +\infty$, we have $\omega_u(\pi)\rightarrow +\infty$. Together with the result that $\omega_u(\pi)$ is an increasing function, there must exist a unique $\pi^*\in[0,+\infty)$ that satisfies $\omega_u(\pi^*)=0$.
\end{proof}

With Proposition 5.1, given WDs' equilibrium response function $\delta^*(\pi)$, the optimal $\pi^*$ as the unique solution to \eqref{uniformopt} can be efficiently obtained via a bi-section search over the feasible price range $\pi^*\in[0,\phi]$, where $\phi$ is a sufficiently large real number.
%Accordingly, we propose an alternating optimization algorithm for the uniform pricing scheme, where the price $\pi$ is optimized iteratively with the common equilibrium threshold $\delta$, until the convergence is achieved.
After obtaining the optimal uniform price in (P2), the remaining optimization of (P2) is a standard Knapsack problem, where off-the-shelf toolboxes can be applied to solve the optimum in pseudo-polynomial time. For example, we can adopt kp01$(\cdot)$ software package in MATLAB \cite{price1}. The details of the proposed algorithm are summarized in Algorithm 1.

\begin{algorithm}[htb]
\caption{Computation of Uniform Pricing and Service Caching for Problem (P2)}
\begin{algorithmic}[1]
\STATE Set $\phi$ as a sufficiently large real number and $\varepsilon=10^{-6}$;
\STATE $\pi^{UB}=\phi$, $\pi^{LB}=0$;
%\STATE The BS initializes the price as $\pi^{(0)}$.
%\STATE Set $t=1$.
\REPEAT
\STATE Set $\pi=\frac{\pi^{UB}+\pi^{LB}}{2}$;
\STATE Calculate $\delta^*(\pi)$ in Stage II according to \eqref{stage2};
\IF{$\omega_u(\pi)<0$}
\STATE $\pi^{LB}=\pi$;
\ELSE
\STATE $\pi^{UB}=\pi$;
\ENDIF
\UNTIL{$\left| \omega_u(\pi) \right|<\varepsilon$.}
%\STATE Apply bi-section search method to find $\delta^{(t-1)}$ that satisfies $\Phi(\delta^{(t-1)})=0$ with given $\pi^{(t-1)}$;
%\STATE Apply bi-section search method to find $\pi^{(t)}$ that satisfies $\omega_u(\pi^{(t)})=0$ with given $\delta^{(t-1)}$;
%\STATE Set $t=t+1$;
%\UNTIL{$\pi^*$ converges.}
\STATE Obtain the optimal caching decisions $\mathbf{x}^*$ by solving knapsack problem via toolbox and using obtained $\pi^*$.
\end{algorithmic}
\end{algorithm}

\subsection{Uniform Distribution of $\theta_i$ in \eqref{definetheta}}

To further obtain some engineering insights from our two-stage game analysis, we consider a special case where $\theta_i$ follows a uniform distribution within $[\underline{\theta},\bar{\theta}]$. Without loss of generality, we assume that $\bar{\theta}>\frac{1}{f^c}$. Otherwise, none of the WDs will offload its task to the BS.  Besides, we generally assume that $\underline{\theta}<0$, implying that $\tau_i^l<\tau_i^u$ for some WDs facing bad channel conditions.

\emph{\textbf{Proposition 5.2:}}  When $\theta_i$ is uniformly distributed in the range $[\underline{\theta},\bar{\theta}]$, the optimal uniform price is
\begin{align}\label{prop4.5}
\pi^*=\frac{\bar{\theta}}{2}-\frac{1}{2f^c}.
\end{align}

\begin{proof}
Based on Theorem 1, we first derive the closed-form expression of $\delta(\pi)$, i.e.,
\begin{align}\label{uniformD}
\delta(\pi)=\left\{
      \begin{array}{ll}
        \frac{M}{f^c}, & \pi<\underline{\theta}-\frac{M}{f^c}; \\
        \frac{(M-1)(\bar{\theta}-\pi)+(\bar{\theta}-\underline{\theta})}{(\bar{\theta}-\underline{\theta})f^c+(M-1)}, & \underline{\theta}-\frac{M}{f^c}\leq\pi\leq\bar{\theta}-\frac{1}{f^c};\\
       \frac{1}{f^c}, & \pi>\bar{\theta}-\frac{1}{f^c}.
      \end{array}
    \right.
\end{align}
By substituting \eqref{uniformD} into \eqref{uniformopt}, we have $\pi^*$ in \eqref{prop4.5}.
%%
%\begin{align}\nonumber
%\pi^*=\frac{\bar{\theta}f^c-\pi^*f^c-1}{(\bar{\theta}-\underline{\theta})f^c+(M-1)}[(\bar{\theta}-\underline{\theta})+\frac{M-1}{f^c}]=\bar{\theta}-\pi^*-\frac{1}{f^c}.
%\end{align}
%%
%That is, $\pi^*=\frac{\bar{\theta}}{2}-\frac{1}{2f^c}$.
\end{proof}

From Proposition 5.2, we have the following interesting observations:
\begin{itemize}
  \item The BS tends to set a higher uniform price when it has more computation power, i.e., a larger $f^c$. This is because a larger $f^c$ increases the WDs' willingness to offload. Thus, the BS can charge a higher price to obtain higher profit.
%Besides, the result in \eqref{prop4.5} is also consistent with the Proposition 4.3 that the price is unrelated to $M$.
  \item Increasing $\bar{\theta}$, the upper bound of the $\theta_i$, leads to a higher optimal price $\pi^*$. It is because a larger average $\theta_i$ (i.e., a larger difference between local execution time $\tau_i^l$ and offloading transmission delay $\tau_i^u$ according to \eqref{definetheta}) represents higher probability to offload for WD $i$. Accordingly, the BS can increase the price for higher profit.

\end{itemize}

Notice that in our proposed uniform pricing scheme, the BS first caches the optimal set of programs and then announces the uniform price for all the programs.
%\footnote{Note that in uniform pricing, we approximate the equilibrium parameter $\delta^*(\pi)$ in Stage II assuming that all the WDs are only informed the uniform price. Because of the limited caching space, many computation tasks with uncached programs may still be offloaded to the edge server in response to the announced uniform price, which not only creates no profit for the BS but also reduces the offloading probabilities of the cached programs. This problem can be addressed by the differentiated pricing scheme in the next section. In particular, the BS sets sufficiently large price for those uncached programs to prohibit the WDs from offloading the corresponding computation tasks.}
Nevertheless, due to the variation in the programs' properties (i.e., the programs' popularity and workloads, the storage sizes of programs and the cost of acquiring different programs), the BS can set different prices for different programs to increase its profit. We will investigate the differentiated pricing scheme in the following section.

\section{Differentiated Pricing to Cope with MEC Service Caching}

\subsection{BS' Profit Maximization in Problem (P1)}

In this section, we consider the general case of Problem (P1) where the BS is allowed to charge different prices for computing different types of tasks.

Based on Stage II's Bayesian equilibrium derived in Theorem 1, the offloading probability of type-$j$ tasks is equal to $1-F(\delta^*(\bm{\pi})+\pi_j)$. Then, Problem (P1) in Stage I is expressed as
\begin{eqnarray}
\mbox{(P3)}~~\max_{(\bm{\pi},\mathbf{x})}&&\sum_{j=1}^N(1-F(\delta^*(\bm{\pi})+\pi_j))x_jq_jM\pi_jL_j-\sum_{j=1}^Nx_jr_j,\nonumber\\
{\rm s.t.}&&\sum_{j=1}^{N}x_jc_j\leq C,\nonumber\\
&&x_{j}\in\{0,1\},\forall j=1,...,N.\label{P3}
\end{eqnarray}
%

%For the above Problem (P5), we can apply the heuristic search methods such as partical swarm optimization (PSO) to find the optimal solution.

Problem (P3) is challenging due to the strong coupling between the caching decisions and each program's task price. Unlike in (P2), the program prices cannot be separately optimized without considering the caching decisions.  To tackle this problem, we first suppose that the caching decisions are given and derive some important properties of the optimal prices in the following proposition, based on which we propose an efficient algorithm to optimize the prices.

\emph{\textbf{Proposition 6.1:}} Suppose that a subset $\hat{\mathcal{N}}\subseteq \mathcal{N}$ of programs are cached in the BS. Then, the optimal program price $\pi_j^*>0, \forall j\in\hat{\mathcal{N}}$ satisfies
\begin{small}
\begin{align}\label{wd}
\omega_{d}(\pi_j):=&q_jML_j\bigg\{(f^c+(M-1)\sum_{k\neq j,k\in\hat{\mathcal{N}}}q_{k}f(\delta^*(\bm{\pi})+\pi_{k}))\left[\frac{[1-F(\delta^*(\bm{\pi})+\pi_j)]}{f(\delta^*(\bm{\pi})+\pi_j)}-\pi_j\right]\nonumber\\
&+(M-1)q_j[1-F(\delta^*(\bm{\pi})+\pi_j)]\bigg\}+(M-1)q_j\sum_{k\neq j,k\in \hat{\mathcal{N}}}f(\delta^*(\bm{\pi})+\pi_{k})q_{k}M\pi_{k}L_{k}=0, \forall j\in\hat{\mathcal{N}}.
\end{align}
\end{small}
Here, $\delta^*(\bm{\pi})$ in Stage II is obtained in \eqref{stage2}.

%Given any other cached programs' price $\pi_{k}, k\in \hat{\mathcal{N}}\setminus j$ and the common equilibrium threshold $\delta(\bm{\pi})$ in \eqref{stage2} in stage II, the optimal price $\pi_j^*>0$ of program $j\in\hat{\mathcal{N}}$ exists and is unique.

\begin{proof}
The expected profit of the BS is
\begin{align}\nonumber
U_B=\sum_{j\in \hat{\mathcal{N}}}(1-F(\delta^*+\pi_j))q_jM\pi_jL_j-\sum_{j\in \hat{\mathcal{N}}}r_j.
\end{align}

To find the optimal price of program $j$, we calculate the derivative of $U_B$ with respect to $\pi_j$ as
\begin{align}\nonumber
\frac{\partial U_B}{\partial\pi_j}=&-f(\delta^*+\pi_j)(1+\frac{\partial\delta^*}{\partial\pi_j})q_jM\pi_jL_j+[1-F(\delta^*+\pi_j)]q_jML_j\\
&+\sum_{k\neq j,k\in \hat{\mathcal{N}}}-f(\delta^*+\pi_{k})\frac{\partial\delta^*}{\partial\pi_j}q_{k}M\pi_{k}L_{k}.\nonumber
\end{align}
According to the analysis in Stage II, we have
\begin{align}\nonumber
\frac{\partial\delta^*}{\partial\pi_j}=\frac{-(M-1)q_jf(\delta^*+\pi_j)}{f^c+(M-1)\sum_{j\in\hat{\mathcal{N}}}q_jf(\delta^*+\pi_j)}.
\end{align}
By equating $\frac{\partial U_B}{\partial\pi_j}=0$, we have
\begin{align}\label{derivativeubpi}
&-(1+\frac{-(M-1)q_jf(\delta^*+\pi_j)}{f^c+(M-1)\sum_{j\in\hat{\mathcal{N}}}q_jf(\delta^*+\pi_j)})q_jM\pi_jL_j+\frac{[1-F(\delta^*+\pi_j)]}{f(\delta^*+\pi_j)}q_jML_j\nonumber\\
&=\sum_{k\neq j,k\in \hat{\mathcal{N}}}f(\delta^*+\pi_{k})\frac{-(M-1)q_j}{f^c+(M-1)\sum_{j\in\hat{\mathcal{N}}}q_jf(\delta^*+\pi_j)}q_{k}M\pi_{k}L_{k}, \forall j\in \hat{\mathcal{N}}.
\end{align}
By multiplying $[f^c+(M-1)\sum_{j\in\hat{\mathcal{N}}}q_jf(\delta^*+\pi_j)]$ in both sides of \eqref{derivativeubpi} for program $j$, we have
\begin{small}
\begin{align}
&q_jML_j\bigg[[f^c+(M-1)\sum_{j\in\hat{\mathcal{N}}}q_jf(\delta^*+\pi_j)]\frac{[1-F(\delta^*+\pi_j)]}{f(\delta^*+\pi_j)}-(f^c+(M-1)\sum_{k\neq j,k\in\hat{\mathcal{N}}}q_{k}f(\delta^*+\pi_{k}))\pi_j\bigg]\nonumber\\
&=q_jML_j\bigg\{(f^c+(M-1)\sum_{k\neq j,k\in\hat{\mathcal{N}}}q_{k}f(\delta^*+\pi_{k}))\left[\frac{[1-F(\delta^*+\pi_j)]}{f(\delta^*+\pi_j)}-\pi_j\right]+(M-1)q_j[1-F(\delta^*+\pi_j)]\bigg\}\nonumber\\
&=-(M-1)q_j\sum_{k\neq j,k\in \hat{\mathcal{N}}}f(\delta^*+\pi_{k})q_{k}M\pi_{k}L_{k},\nonumber
\end{align}
\end{small}
which yields $\omega_{d}(\pi_j)=0$.
%Then, we define
%%
%\begin{small}
%\begin{align}\label{wd}
%\omega_{d}(\pi_j)=&q_jML_j\bigg\{(f^c+(M-1)\sum_{k\neq j,k\in\hat{\mathcal{N}}}q_{k}f(\delta+\pi_{k}))\left[\frac{[1-F(\delta+\pi_j)]}{f(\delta+\pi_j)}-\pi_j\right]+(M-1)q_j[1-F(\delta+\pi_j)]\bigg\}\nonumber\\
%&+(M-1)q_j\sum_{k\neq j,k\in \hat{\mathcal{N}}}f(\delta+\pi_{k})q_{k}M\pi_{k}L_{k},
%\end{align}
%\end{small}
%%
%and the optimal price $\pi_j^*$ satisfies $\omega_{d}(\pi_j^*)=0$.
\end{proof}

Accordingly, we propose an alternating algorithm that alternately optimizes the prices of the cached programs given any feasible $\hat{\mathcal{N}}$. Specifically, in the $t$-th iteration, the algorithm finds the optimal $\{\pi^{(t)}_j,j\in\hat{\mathcal{N}}\}$ according to Proposition 6.1 given $\bm{\pi}_{-j}^{(t-1)}=\{\pi_k^{(t-1)}, k\in\hat{\mathcal{N}}\setminus j\}$ and the corresponding $\delta^*(\bm{\pi}^{(t-1)})$. In the following, we show that there exists a unique $\{\pi^{(t)}_j,j\in\hat{\mathcal{N}}\}$ in each iteration $t$.

\emph{\textbf{Corollary 6.1:}} Suppose that the other prices $\bm{\pi}_{-j}^{(t-1)}$ and $\delta^*(\bm{\pi}^{(t-1)})$ in the last iteration are given. There exists a unique solution $\pi^{(t)}_j, j\in\hat{\mathcal{N}},$ that satisfies $\omega_{d}(\pi^{(t)}_j)=0$ in the $t$-th iteration.

\begin{proof}
%Next, we show that there exists a unique $\pi_j^*$ satisfying $\omega_{d}(\pi_j^*)=0$ when $\pi_{-j}$ and $\delta^*$ are given.
If the distribution of $\theta_i$ is regular, then $\frac{[1-F(\delta^*(\bm{\pi}^{(t-1)})+\pi_j)]}{f(\delta^*(\bm{\pi}^{(t-1)})+\pi_j)}-\pi_j$ is a decreasing function in $\pi_j$ given $\delta^*(\bm{\pi}^{(t-1)})$. Besides, $(M-1)q_j[1-F(\delta^*(\bm{\pi}^{(t-1)})+\pi_j)]$ decreases in $\pi_j$. Therefore, $\omega_d(\pi_j)$ is a decreasing function of $\pi_j$ given $\bm{\pi}_{-j}^{(t-1)}$ and  $\delta^*(\bm{\pi}^{(t-1)})$.

When $\pi_j=0$, we have $\omega_{d}(\pi_j)>0$. Besides, when $\pi_j\rightarrow +\infty$, $\omega_{d}(\pi_j)\rightarrow -\infty$. Hence, there exists a unique $\pi_j^*\in(0,+\infty)$ that satisfies $\omega_{d}(\pi_j^*)=0$ given $\bm{\pi}_{-j}^{(t-1)}$ and $\delta^*(\bm{\pi}^{(t-1)})$.
\end{proof}

According to Corollary 6.1, given the prices $\bm{\pi}_{-j}^{(t-1)}$ and the equilibrium parameter $\delta^*(\bm{\pi}^{(t-1)})$ in \eqref{stage2} in Stage II, we can obtain the optimal price $\pi_j^{(t)}$ for the cached program $j, j\in\hat{\mathcal{N}},$ by an efficient bi-section search method over $\pi_j^{(t)}\in[0,\phi]$ that satisfies $\omega_d(\pi_j^{(t)})=0$ in the $t$-th iteration. We summarize the proposed alternating algorithm for differentiated pricing given any feasible $\hat{\mathcal{N}}$ in Algorithm 2.

Then, the remaining optimization of Problem (P3) is to find the optimal caching decisions $\mathbf{x}$. In general, when the total number of programs is moderate, we can enumerate all feasible service caching decisions $\mathbf{x}$ that satisfy the caching space constraints in (P3) and choose the best service caching decision that yields the maximal $U_B$. When the dimension of $\mathbf{x}$ is high, many meta-heuristic methods, such as Gibbs sampling \cite{myfirstTWC} and particle swarm optimization \cite{PSO}, can be applied to effectively find the optimal solution.

\begin{algorithm}[h]
\caption{Computation of Differentiated Pricing given a Feasible $\hat{\mathcal{N}}$ for Problem (P3)}
\begin{algorithmic}[1]
\STATE The BS initializes the prices as $\{\pi_j^{(0)}, j\in\hat{\mathcal{N}}\}$ and calculates the $\delta^*(\bm{\pi}^{(0)})$ in \eqref{stage2};
\STATE Set $t=1$;
\REPEAT
%\STATE Apply bi-section search method to find $\delta^{(t-1)}$ that satisfies $\Phi(\delta^{(t-1)})=0$ in \eqref{stage2} with given $\{\pi_j^{(t-1)}, j\in\hat{\mathcal{N}}\}$;
\FOR{Each cached program $j\in\hat{\mathcal{N}}$}
\STATE Apply bi-section search method to find $\pi_j^{(t)}$ that satisfies $\omega_d(\pi_j^{(t)})=0$ in \eqref{wd} with given $\bm{\pi}_{-j}^{(t-1)}$ and $\delta^*(\bm{\pi}^{(t-1)})$;
%\REPEAT
%\STATE Set $\pi_j^{(t)}=\frac{\pi^{UB}+\pi^{LB}}{2}$;
%\IF{$\omega_d(\pi_j^{(t)}|\{\pi_{j'}^{(t-1)},j'\in\mathcal{N}'/j\},\delta^{(t-1)})>0$}
%\STATE $\pi^{LB}=\pi_j^{(t)}$;
%\ELSE
%\STATE $\pi^{UB}=\pi_j^{(t)}$;
%\ENDIF
%\UNTIL{$\left| \omega_d(\pi_j^{(t)}|\{\pi_{j'}^{(t-1)},j'\in\mathcal{N}'/j\},\delta^{(t-1)}) \right|<\varepsilon$.}
\ENDFOR
\STATE Obtain $\{\pi_j^{(t)}, j\in\hat{\mathcal{N}}\}$ and calculate the corresponding $\delta^*(\bm{\pi}^{(t)})$ in \eqref{stage2};
\STATE Set $t=t+1$.
\UNTIL{$\{\pi_j^*, j\in\hat{\mathcal{N}}\}$ converges.}
\end{algorithmic}
\end{algorithm}

Though complicated, we manage to derive some interesting properties of the optimal prices in the differentiated pricing scheme under special cases to deliver more engineering insights.

\emph{\textbf{Proposition 6.2:}} If the task workloads of all the cached programs are the same ($L_j=L_{k}, \forall j,k \in \hat{\mathcal{N}}$), then the edge server should set the same price ($\pi^*_j = \pi^*,  \forall j \in \hat{\mathcal{N}}$) for all the cached programs as the unique solution to
\begin{align}\label{diffoptcon}
\pi^*-\frac{1-F(\delta^*(\pi^*)+\pi^*)}{f(\delta^*(\pi^*)+\pi^*)}-\frac{(M-1)(\sum_{j\in\hat{\mathcal{N}}}q_j)(1-F(\delta^*(\pi^*)+\pi^*))}{f^c}=0.
\end{align}
%
%where $\zeta'=f^c+(M-1)\sum_{j\in \mathcal{N}/\mathcal{N}'}q_jf(\delta^*+\pi_j^*)$.
\begin{proof}
Please refer to Appendix \ref{appendicesG}.
\end{proof}

The above proposition indicates that if the cached programs have the same workload, then the optimal prices per CPU cycle are equal for these programs.

Moreover, we have the following corollary when $f^c\rightarrow\infty$.

\emph{\textbf{Corollary 6.2:}} In the case where the BS has very large computational power, i.e., $f^c\rightarrow\infty$, we have
\begin{align}\nonumber
\pi_j^*=\frac{1-F(\pi_j^*)}{f(\pi_j^*)}, \forall j\in\hat{\mathcal{N}}.
\end{align}

\begin{proof}
When $f^c\rightarrow\infty$, we have $\delta^*=0$ at the Stage II according to Theorem 1. Then,
%for the uniform pricing scheme, based on the Proposition 4.1, when $f^c\rightarrow\infty$, we have $\pi^*=\frac{\left[1-F(\pi^*)\right]}{f(\pi^*)}$.For the differentiated pricing scheme,
according to Proposition 6.1, we have $\pi_j^*=\frac{1-F(\pi_j^*)}{f(\pi_j^*)}, \forall j\in\hat{\mathcal{N}}$.
\end{proof}

From Corollary 6.2, the optimal prices only depend on the PDF and CDF of the WDs' valuation of $\theta_i$ when the $f^c$ is very large. It is intuitive as the WDs do not compete for the computation resource of the BS when it is abundant. Notice that even if $f^c$ goes to infinity, some WDs still choose not to offload due to the long communication delay (i.e., $\tau_i^u>\tau_i^l$ in \eqref{definetheta}).

\subsection{Uniform Distribution of $\theta_i$ in \eqref{definetheta}}

%In order to demonstrate more insights into the interactions between the optimal prices and programs' individual  properties (i.e., workloads $L_j$ and popularity $q_j$) in the differentiated pricing scheme,
Similar as Section V-B, we derive some interesting properties of the optimal prices when $\theta_i$ follows a uniform distribution.

\emph{\textbf{Proposition 6.3:}} Suppose that there are two programs $j$ and $k$ cached at the BS. Then, with uniform distribution of $\theta_i$, we have
\begin{itemize}
  \item If $L_{j}=L_{k}$, then $\pi_{j}^*=\pi_{k}^*=\frac{\bar{\theta}}{2}-\frac{1}{2f^c}$;
  \item If $L_{j}>L_{k}$, then $\pi_{j}^*<\pi_{k}^*$, implying that a higher unit price is charged to the program with the smaller workload.
  %\item If $L_{j}<L_{j'}$, $\pi_{j}^*>\pi_{j'}^*$.
\end{itemize}

\begin{proof}
Please refer to Appendix \ref{appendicesE}.
\end{proof}

From the above proposition, we have the following observations:
\begin{itemize}
  \item If the workloads of different cached programs are equal, the BS sets the same price per CPU cycle for the programs, which is in consistence with Proposition 6.2.
  \item If the workloads are different, the program with the larger workload has a lower price. Notice that the programs' prices are charged for unit computation workload. Intuitively, the BS can increase its profit by setting a lower price for the program with the larger workload to encourage more workloads offloaded from the WDs.
  %\item The above results are unrelated to the program popularity.
\end{itemize}

Besides the relation between the optimal prices and the program workloads, Proposition 6.4 further discusses the impact of program popularity $q_j$ on the optimal prices.

\emph{\textbf{Proposition 6.4:}} Suppose that the BS caches two programs $j$ and $k$. With uniform distribution of $\theta_i$, we have the following properties:
\begin{itemize}
  \item If $L_{j}>L_{k}$, $\pi_j^*$ increases with popularity $q_j$ of program $j$, regardless of the value of $q_k$.
  \item If $L_{j}<L_{k}$, $\pi_j^*$ decreases with $q_j$ when $q_{k}\in (0,\frac{2f^c(\bar{\theta}-\underline{\theta})L_j}{(L_{k}-L_{j})(M-1)})$. Otherwise, when $q_k\geq\frac{2f^c(\bar{\theta}-\underline{\theta})L_j}{(L_{k}-L_{j})(M-1)}$, $\pi_j^*$ increases with $q_j$.
\end{itemize}

\begin{proof}
Please refer to Appendix \ref{appendicesF}.
\end{proof}

From Proposition 6.4, we obtain the following insights:
\begin{itemize}
  \item The optimal price $\pi_j$ increases with $q_j$ if the program $j$ has the larger workload.
%According to Proposition 6.3, the BS focuses on incentivizing offloading of the type-$j$ tasks with larger workload by deciding a lower price compared with the price of the program with smaller workload.
As the program $j$'s popularity increases, more WDs are interested in program $j$ with more demands. Thus, the BS can charge higher price for higher profit.
%needs to decide a lower price to incentivize offloading of type-$j$ tasks with larger workload. In this case, when more WDs are interested in the program $j$ with larger workload, the BS can increase price $\pi_j$ and obtain a larger expected revenue.
  \item If program $j$ has a smaller workload, the relation between $\pi_j$ and $q_j$ depends on the other larger workload's program popularity $q_k$. Specifically, when $q_k$ is small, i.e., the $k$-th program is not popular, the BS has more spare computation resource to serve type-$j$ tasks. Thus, when $q_j$ increases, the BS has the incentive to decrease the price $\pi_j$ to encourage the offloading of more type-$j$ tasks. On the other hand, when $q_k$ is large, the BS has the incentive to discourage the type-$j$ tasks' offloading by setting a higher price $\pi_j$ with the increase of $q_j$, so that it can spare more computation resource for the type-$k$ tasks.
\end{itemize}

\section{Simulation Results}

In this section, we conduct numerical simulations to evaluate the performances of our proposed two-stage dynamic game of incomplete information for service caching, pricing, and task offloading in MEC systems.  For simplicity of illustration, we normalize the unit cost for acquiring each program as $r_j=1$. Besides, we assume that the program sizes $c_j$ are equal for all programs $j$, such that the caching space $C$ at the BS can easily tell the number of cached programs.  $\theta_i$ is assumed to follow a uniform distribution between $-10\times 10^{-8}$ and $10\times 10^{-8}$.

\subsection{WDs' Task Offloading Behaviors in Stage II}
We first investigate the WDs' task offloading behaviors in Stage II. For illustration purpose, we consider a two-program case (i.e., $N=2$) in the following. We assume that there are $M=100$ WDs and the computation capability at the BS $f^c=10^8$ cycles/second.

%We first validate Proposition 4.1 in Section IV.
Fig. 3 illustrates the impact of different programs' prices on the offloading probabilities of two types of tasks, where the program popularity $\{q_j\}=[0.5,~0.5]$. Given the price of program $2$ in the first subfigure of Fig. 3, the offloading probability of type-$1$ tasks decreases with $\pi_1$. Besides, increasing $\pi_2$ leads to a higher offloading probability of type-$1$ tasks. It is due to the fact that increasing $\pi_2$ reduces the offloading probability of type-$2$ tasks, which makes more computation resource available for computing the type-$1$ tasks. Nevertheless, the offloading probability of type-$2$ tasks in the second subfigure of Fig. 3 shows an opposite trend, where the offloading probability increases with $\pi_1$ and decreases with $\pi_2$. An interesting observation is that when $\pi_1$ is high\footnote{In this paper, the program prices are charged for unit CPU cycle. The total payment $\pi_1L_1$ from the WD with type-1 task is nontrivial.} (e.g., above $4\times 10^{-8}$ when $\pi_2=2\times 10^{-8}$), the offloading probability of type-$2$ tasks is fixed. It is because the offloading probability of type-$1$ tasks drops to zero as shown in the left subfigure of Fig. 3. In this case, further increasing the price of program $1$ does not affect the offloading probability of type-$2$ tasks.

\begin{figure*}[htbp]
\centering
\begin{minipage}[t]{0.45\textwidth}
\centering
\includegraphics[scale=0.5]{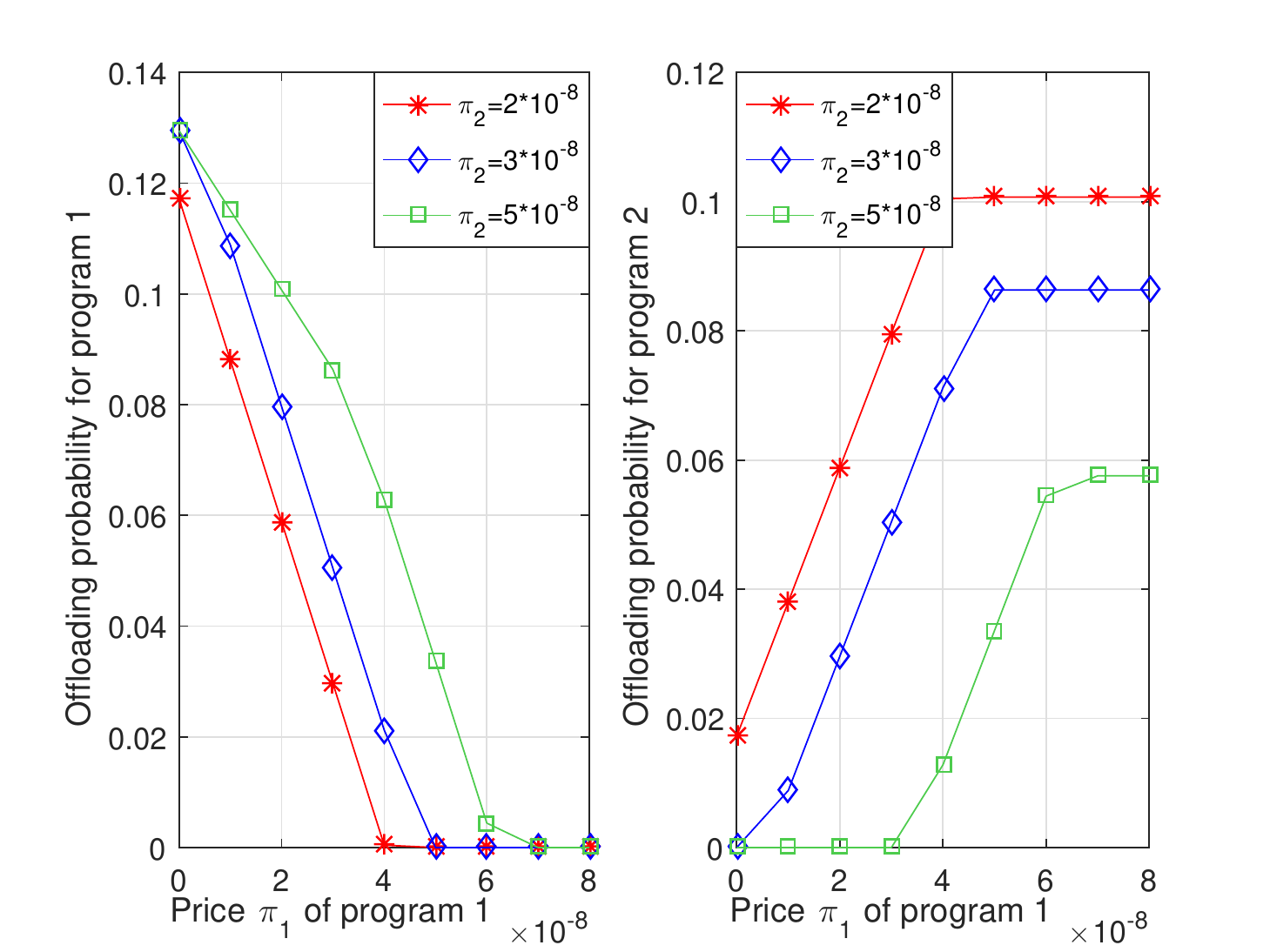}
\caption{Offloading probability as a function of $\pi_1$ under different $\pi_2$.}
\end{minipage}
\begin{minipage}[t]{0.45\textwidth}
\centering
\includegraphics[scale=0.5]{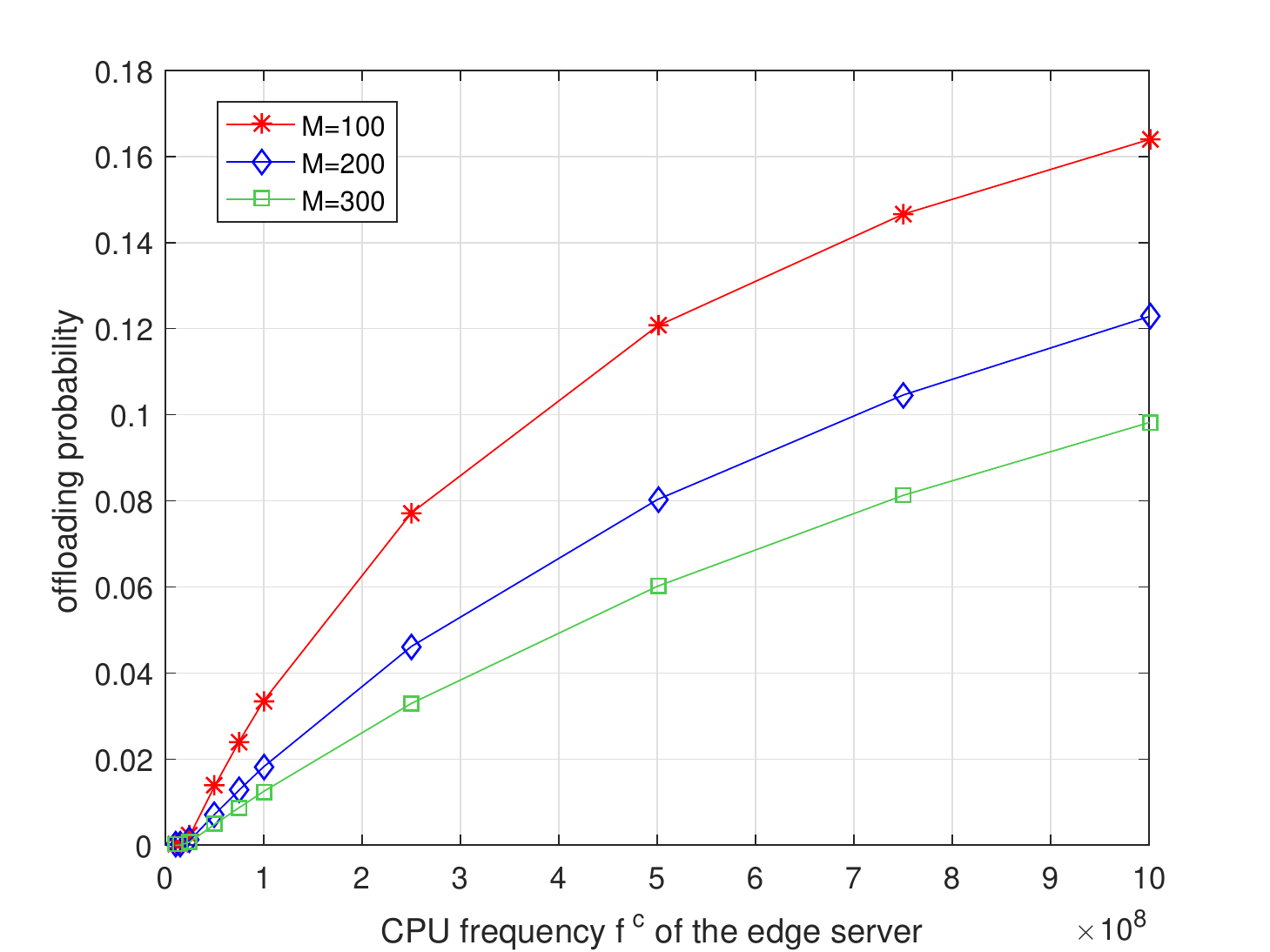}
\caption{Offloading probability as a function of $f^c$  under different numbers of WDs $M$.}
\end{minipage}
\end{figure*}

%Then, we verify Proposition 4.2 in Section IV.
Fig. 4 studies the impact of the computation capability on the offloading probability of each program's tasks under different number of WDs $M$, where the program popularity $\{q_j\}=[0.5,~0.5]$. We set equal price $5\times 10^{-8}$ for both programs, which leads to the same offloading probabilities of different types of tasks.  We observe that the WDs' offloading probability under each program increases in the edge server's CPU frequency $f^c$. Besides, increasing $M$ or WDs' competition in sharing $f^c$  reduces the offloading probability. These also coincide with our results in Proposition 4.2.

\subsection{BS' Service Caching and Pricing Strategies in Stage I}
Then, we show the properties of BS' service caching and pricing strategies in Stage I for the considered two-program case, where the BS is able to cache at most two programs, i.e., $C=2$.

\begin{figure*}[htb]
  \centering
\subfigure[Equal workloads of the two programs' tasks  ]{\includegraphics[width=0.45\textwidth]{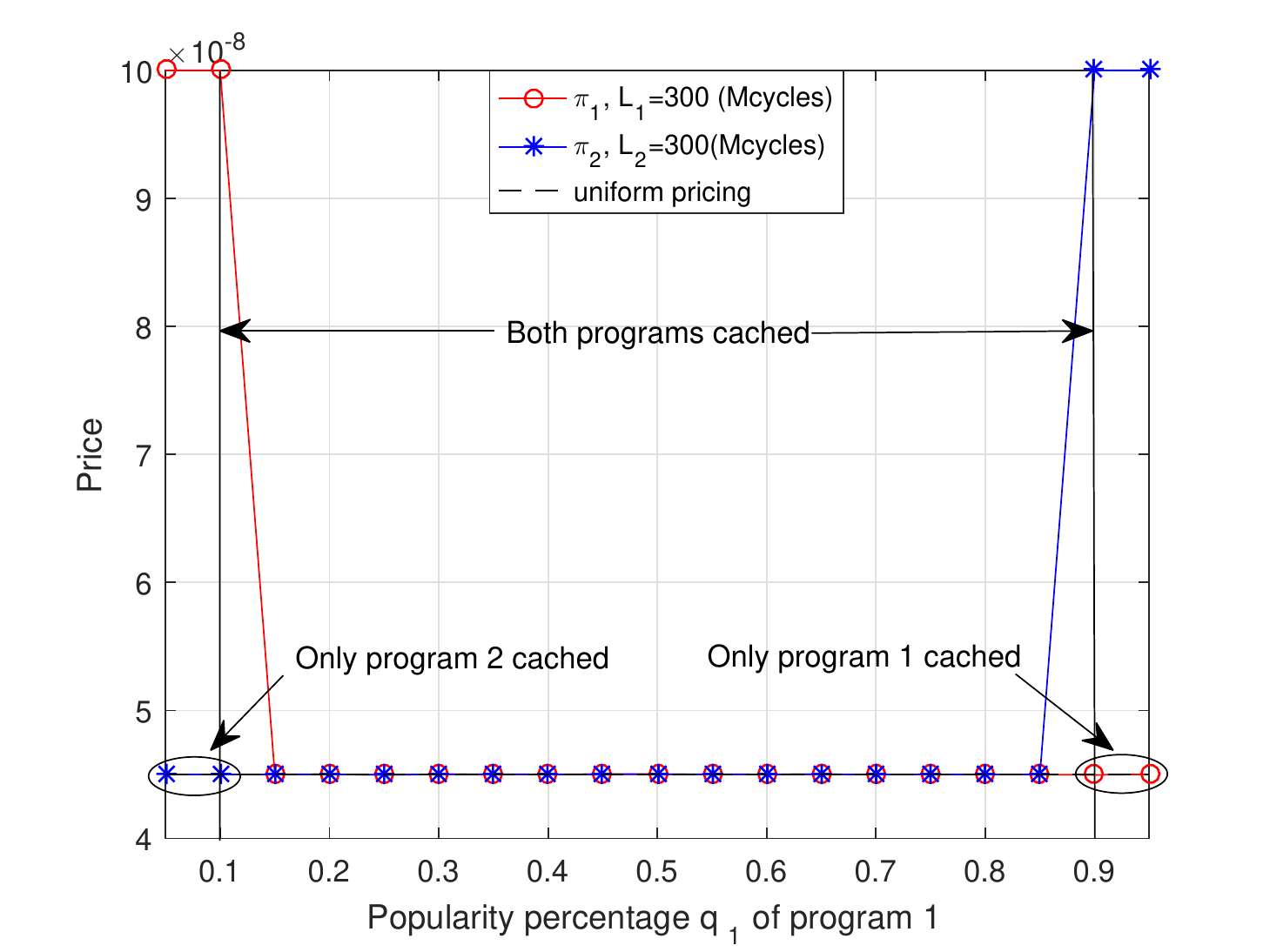}}
    \subfigure[Different workloads of the two programs' tasks  ]{\includegraphics[width=0.45\textwidth]{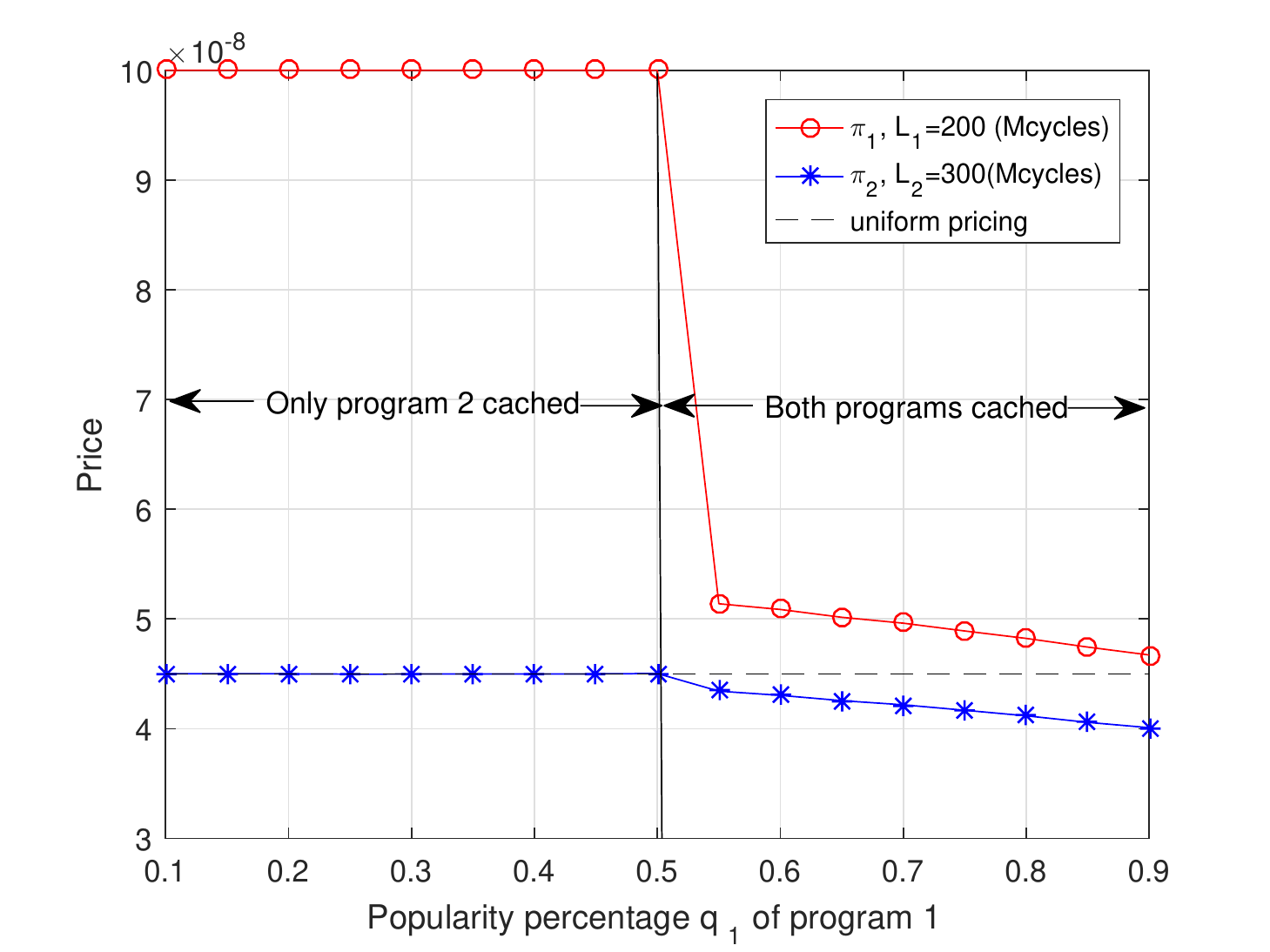}}
     \\
  \caption{The price for each program versus the popularity $q_1$ in the two-program case.}
    % \label{fig:data_distribution}
    %\vspace{0.2in}
\end{figure*}

In Fig. 5, we demonstrate the impact of program popularity on the optimal prices, where $f^c=10^8$ cycles/s and $M=100$. For the uniform pricing scheme, the optimal price is fixed regardless of the program popularity as shown in Proposition 5.1. For the differentiated pricing scheme, when the workloads of two
programs are equal (i.e., $\{L_j\}=[300,~300]$ Mcycles), we observe from Fig. 5(a) that if both programs are cached, the prices of the two programs are the same and equal to the optimal uniform price. When the popularity of program 1 is small (i.e., below 0.1), the BS does not cache program 1 and sets a sufficiently large price (i.e., $\pi_1=10\times 10^{-8}$) for program 1 to guarantee zero offloading probability of type-$1$ tasks. It is because in this case, the BS' profit obtained by caching program 1 cannot compensate for the program 1's acquiring cost charged by the program provider.
Besides, as shown in Fig. 5(b), when the workloads of two programs are different (i.e., $\{L_j\}=[200,~300]$ Mcycles), we can see that the BS always caches program 2 with larger workload. It is due to the fact that higher total revenue $L_2\pi_2$ is obtained for executing one type-2 task with larger workload. As $q_1$ is larger than 0.5, both programs 1 and 2 are cached and the price of program 1 (lower workload) is higher than that of program 2. Besides, we observe that, when both programs are cached, with the increase of $q_1$, both $\pi_1$ and $\pi_2$ decrease. It is because $q_2=1-q_1$ in this case and equivalently, $\pi_2$ increases in $q_2$, which coincides with our analytical results in Proposition 6.4.

\begin{figure*}[htbp]
\centering
\begin{minipage}[t]{0.45\textwidth}
\centering
\includegraphics[scale=0.5]{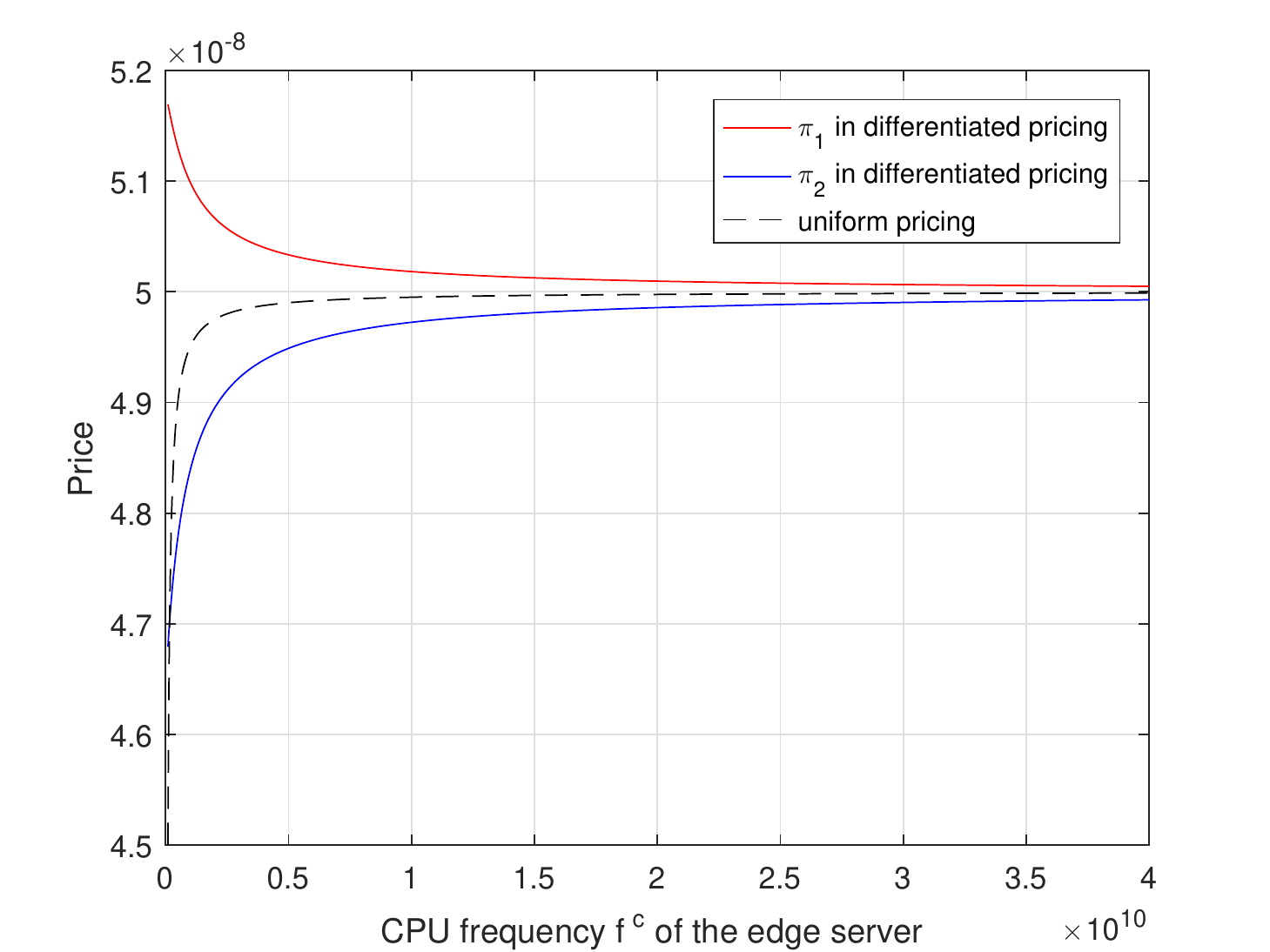}
\caption{The price for each program versus the edge computation capability $f^c$ in the two-program case.}
\end{minipage}
\begin{minipage}[t]{0.45\textwidth}
\centering
\includegraphics[scale=0.5]{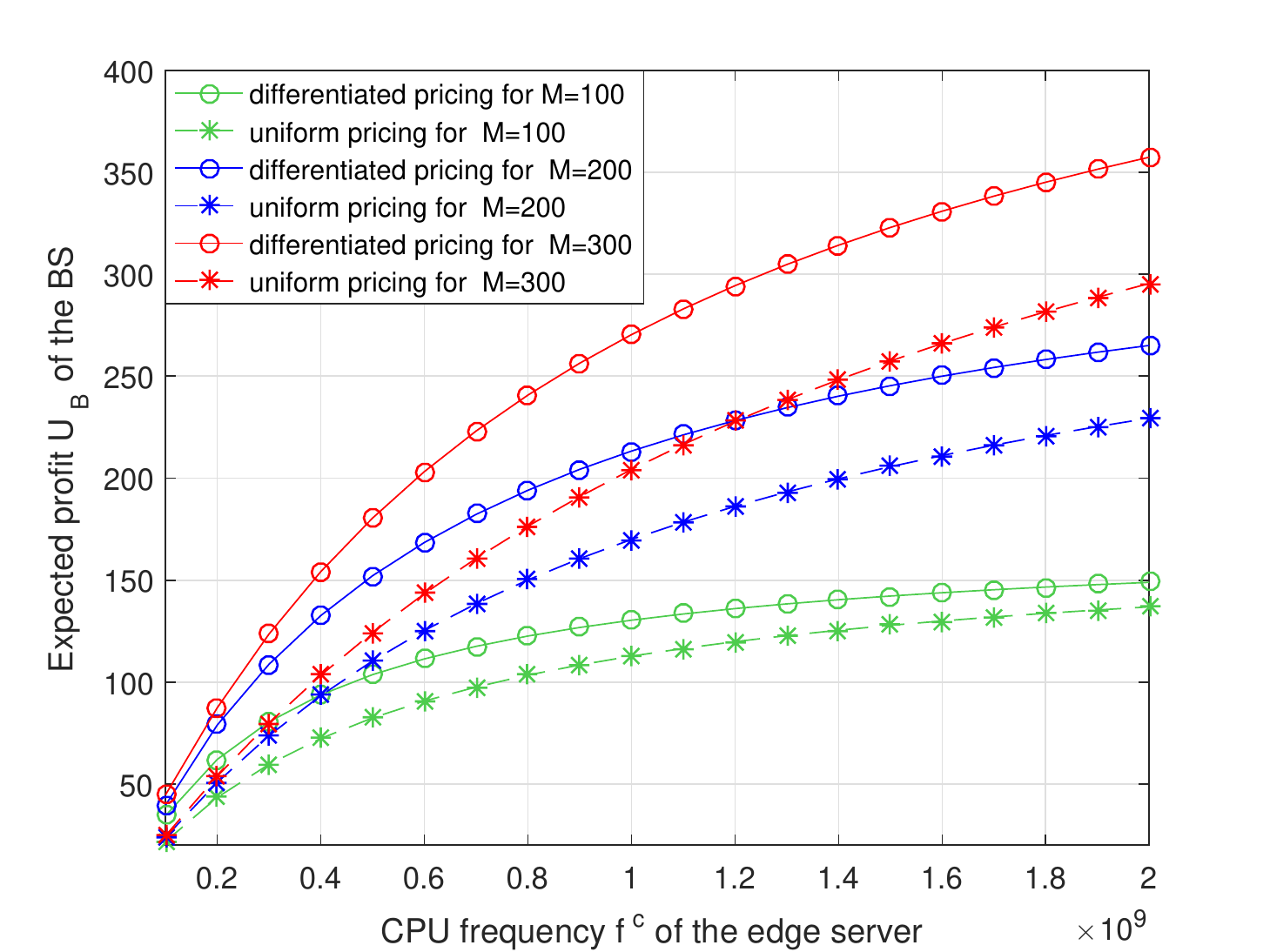}
\caption{The expected profit of the BS versus edge computation capability $f^c$ under different WDs number $M$.}
\end{minipage}
\end{figure*}

Furthermore, Fig. 6 illustrates the impact of edge computation capability $f^c$ on the optimal prices, where $\{L_j\}=[200,~300]$ Mcycles, $\{q_j\}=[0.6,~0.4]$ and $M=100$. For the uniform pricing scheme, the optimal price increases with $f^c$ as analyzed in Proposition 5.2. As for the differentiated pricing scheme, it is observed that when $f^c$ increases, the optimal price of the program 2 with larger workload increases. It is because the BS always sets a lower price for the program with larger workload in order to attract more WDs interested in this program to offload. Accordingly, as $f^c$ increases, the BS can decide a higher price $\pi_2$ and obtain a larger profit. However, the price $\pi_1$ of the program 1 with lower workload decreases in $f^c$ so as to incentivize offloading of type-$1$ tasks since the BS has more spare computation resource. In addition, we find that for a sufficiently large $f^c$, the prices of both programs tend to be the same, which is consistence with Corollary 6.2. Besides, we observe that the uniform price is in between the differentiated prices $\pi_1$ and $\pi_2$, as the uniform pricing scheme is like an average way to coordinate the two cached programs and guide task offloading.

\subsection{Performance Comparison Between Uniform and Differentiated Pricing Algorithms}

Furthermore, we evaluate and compare the performances of the two proposed pricing algorithms in term of gaining MEC service profit for the BS. We assume that the WDs' tasks belong to $N=3$ programs with popularity $\{q_j\}=[0.2,~0.4,~0.4]$. Here, we set $C=2$ for at most two programs to cache. The computing workloads for the programs are $\{L_j\}=[300,~200,~100]$ (Mcycles).

Note that the uniform pricing (though easier and more fair to implement in practice) is a special case of the differentiated pricing, and yields smaller expected profit for the BS. In Fig. 7, we present the expected profit of the BS using the two pricing algorithms versus CPU frequency $f^c$ at the edge server and the number of WDs $M$. We observe that as $f^c$ or $M$ increases, higher expected profit of the BS is obtained under both uniform and differentiated pricing schemes, and uniform pricing still gains most profit of the differentiated pricing for different settings, telling the value to adopt simpler uniform pricing in practice.
The profit gap between differentiated pricing upperbound and uniform pricing increases as there are more WDs with greater $M$ to tailor for each program's WDs.
%Besides, under any number of WDs $M$, we observe that the differentiated pricing scheme outperforms the uniform pricing scheme. Moreover, with the increase of $f^c$, the revenue increasing speed becomes lower, which verifies the Corollary 5.1 that the differentiated pricing strategy only depends on the distribution of $\theta_i$ when the BS has infinite computation capability.

\section{Conclusions}

This paper has studied a pricing mechanism to coordinate service caching and guide task offloading in an MEC system with one BS and multiple associated WDs. We have proposed a two-stage dynamic game of incomplete information to capture the interaction between the BS and WDs. In Stage I, the BS aims to maximize its expected profit by optimizing its service caching decisions and the programs prices for WDs' task executions  under the limited computation resource and caching storage capacity. In Stage II, for given prices of service programs, the WDs play a Bayesian subgame and selfishly optimize offloading decisions to minimize their own costs by estimating the other WDs' decisions. We have first derived the threshold-based offloading strategy among the WDs at the Bayesian equilibrium. Then, by predicting the WDs' offloading equilibrium, in Stage I, we have developed the uniform and differentiated pricing algorithms to optimize the prices and service caching decisions at the BS. Simulation results have validated our analysis and shown the effectiveness of our proposed pricing mechanism.

\appendices

\section{Proof of Proposition 4.1} \label{appendicesA}
According to Theorem 1, we have $\Phi(\pi_j,\delta^*(\pi_j))=0$. By applying implicit function theorem, we have
$\frac{\partial\Phi(\delta^*)}{\partial\pi_j}+\frac{\partial\Phi(\delta^*)}{\partial\delta^*}\frac{\partial\delta^*}{\partial\pi_j}=0, \forall j\in\mathcal{N}$.
That is,
\begin{align}\nonumber
\frac{M-1}{f^c}\bigg[q_j\frac{\partial F(\delta^*+\pi_j)}{\partial(\delta^*+\pi_j)}\bigg]+\frac{\partial\delta^*}{\partial\pi_j}\bigg[1+\frac{M-1}{f^c}\sum_jq_j\frac{\partial F(\delta^*+\pi_j)}{\partial(\delta^*+\pi_j)}\bigg]=0.
\end{align}
Hence,
\begin{align}\nonumber
\frac{\partial\delta^*}{\partial\pi_j}=\frac{-(M-1)q_jf(\delta^*+\pi_j)}{f^c+(M-1)\sum_jq_jf(\delta^*+\pi_j)},\forall j\in\mathcal{N}.
\end{align}
Therefore, for the relation between $\delta^*+\pi_j$ and $\pi_j$, we have
\begin{align}\nonumber
\frac{\partial(\delta^*+\pi_j)}{\partial\pi_j}=1-\frac{(M-1)q_jf(\delta^*+\pi_j)}{f^c+(M-1)\sum_jq_jf(\delta^*+\pi_j)}>0.
\end{align}
Besides, for the relation between $\delta^*+\pi_j$ and $\pi_{k}, k\in\mathcal{N}\setminus j$, we have
\begin{align}\nonumber
\frac{\partial(\delta^*+\pi_j)}{\partial\pi_{k}}=\frac{\partial\delta^*}{\partial\pi_{k}}=\frac{-(M-1)q_{k}f(\delta^*+\pi_{k})}{f^c+(M-1)\sum_jq_jf(\delta^*+\pi_j)}<0.
\end{align}

\section{Proof of Proposition 4.2} \label{appendicesB}

We first prove the relation between $\delta^*$ and $f^c$. Equation \eqref{stage2} shows that $\Phi(\delta^*)$ is increasing in $f^c$. Recall that $\Phi(\delta^*)$  is increasing in $\delta^*$. Based on $\Phi(f^c,\delta^*(f^c))=0$, by applying implicit function theorem, we have
$\frac{\partial\Phi(\delta^*)}{\partial f^c}+\frac{\partial\Phi(\delta^*)}{\partial\delta^*}\frac{\partial\delta^*}{\partial f^c}=0$.
Accordingly, we can derive
$\frac{\partial\delta^*}{\partial f^c}=-\frac{\partial\Phi(\delta^*)}{\partial f^c}/\frac{\partial\Phi(\delta^*)}{\partial\delta^*}<0.$
Thus, $\delta^*$ is decreasing in $f^c$. Accordingly, given the price $\pi_j$, $\delta^*+\pi_j$ also decreases in $f^c$.

Then, we prove the relation between $\delta^*$ and $M$. Equation \eqref{stage2} shows that $\Phi(\delta^*)$ decreases with $M$. Suppose that $M_1<M_2$ and $\Phi(M_1,\delta^*_1)=0$. Because $\Phi(\delta^*)$ is increasing in $\delta^*$, to maintain $\Phi(M_2,\delta^*_2)=0$, $\delta^*_2>\delta^*_1$ must hold. Therefore, $\delta^*$ is increasing in $M$. Accordingly, given the price $\pi_j$, $\delta^*+\pi_j$ also increases in $M$.

\section{Proof of Proposition 6.2}\label{appendicesG}

We prove the existence, uniqueness and optimality of the solution, respectively.

\emph{Existence and uniqueness:} According to \eqref{diffoptcon}, we define
\begin{small}
\begin{align}
\omega_{eq}(\pi)&=\pi-\frac{1-F(\delta(\pi)+\pi)}{f(\delta(\pi)+\pi)}-\frac{(M-1)(\sum_{j\in\hat{\mathcal{N}}}q_j)(1-F(\delta(\pi)+\pi))}{f^c}\nonumber\\
&=\delta(\pi)+\pi-\frac{1-F(\delta(\pi)+\pi)}{f(\delta(\pi)+\pi)}-\frac{2(M-1)(\sum_{j\in\hat{\mathcal{N}}}q_j)(1-F(\delta(\pi)+\pi))+1}{f^c}.
\end{align}
\end{small}
%
%%
%\begin{align}
%\omega_{eq}(\pi)=(1-F(\delta+\pi))-\frac{f^c}{f^c+(M-1)(\sum_{j\in\mathcal{N}'}q_j)f(\delta+\pi)}f(\delta+\pi)\pi.
%\end{align}
%%
According to Proposition 4.1 in Stage II, $\delta(\pi)+\pi$ increases in $\pi$. Therefore, under regular distribution assumption, $\delta(\pi)+\pi-\frac{1-F(\delta(\pi)+\pi)}{f(\delta(\pi)+\pi)}$ is an increasing function with respect to $\pi$. Besides, $-\frac{2(M-1)(\sum_{j\in\hat{\mathcal{N}}}q_j)(1-F(\delta(\pi)+\pi))+1}{f^c}$ also increases in $\pi$. Hence, all terms in $\omega_{eq}(\pi)$ increase in $\pi$, thus  $\omega_{eq}(\pi)$ is an increasing function in $\pi$. Meanwhile, when $\pi=0$, we have $\omega_{eq}<0$. When $\pi\rightarrow +\infty$, $\omega_{eq}\rightarrow +\infty$. Thus, there exists a unique $\pi^*>0$ that satisfies $\omega_{eq}(\pi^*)=0$.

\emph{Optimality:} By substituting \eqref{diffoptcon} into $\omega_{d}(\pi_j)$ for each cached program $j$, we have
\begin{small}
\begin{align}
\omega_{d}(\pi_j)&=q_jML_j\bigg[[f^c+(M-1)\sum_{j\in\hat{\mathcal{N}}}q_jf(\delta^*+\pi_j)]\frac{f^c}{f^c+(M-1)\sum_{j\in\hat{\mathcal{N}}}q_jf(\delta^*+\pi_j)}\pi_j\nonumber\\
&-(f^c+(M-1)\sum_{k\neq j,k\in\hat{\mathcal{N}}}q_{k}f(\delta^*+\pi_{k}))\pi_j\bigg]+(M-1)q_j\sum_{k\neq j,k\in \hat{\mathcal{N}}}f(\delta^*+\pi_{k})q_{k}M\pi_{k}L_{k}\nonumber\\
&=-(M-1)q_j\sum_{k\neq j,k\in\hat{\mathcal{N}}}q_{k}f(\delta^*+\pi_{k})M\pi_jL_j+(M-1)q_j\sum_{k\neq j,k\in \hat{\mathcal{N}}}q_{k}f(\delta^*+\pi_{k})M\pi_{k}L_{k}, \forall j\in \hat{\mathcal{N}}. \nonumber
\end{align}
\end{small}
Since the price and workload for each program are equal, we have $L_1\pi_1^*=L_2\pi_2^*=...=L_j\pi_j^*, \forall j\in \hat{\mathcal{N}}$.
Therefore, we have $\omega_{d}(\pi_j^*)=0,\forall j\in \hat{\mathcal{N}}$. It completes the proof.
%\begin{small}
%\begin{align}
%LHS=-\sum_{j'\neq j,j'\in \mathcal{N}'}(M-1)q_{j'}f(\delta^*+\pi_{j'})\pi_{j'}L_{j'}, \forall j\in \mathcal{N}', \nonumber
%\end{align}
%\end{small}
%%
%which is equal to the RHS of \eqref{derivativeubpi} for each cached program $j$.

%%%%%%%%%%%%%%%%%%%%%%%%%%%%%%%%%%%%%%%%%%%%%%%%%%%%%%%%%%%%%%%%%%%%%%%%%%%%%%%%%%%%%%%%%%%%%%%%%%%%%%%%%%%%%%

\section{Proof of Proposition 6.3} \label{appendicesE}
Suppose that there are two programs $j$ and $k$ cached in the BS with popularity $q_{j}$ and $q_{k}$. From \eqref{P3}, the expected profit of the BS is
\begin{small}
\begin{align}\nonumber
U_B=\frac{\bar{\theta}-(\delta^*+\pi_j)}{\bar{\theta}-\underline{\theta}}q_jM\pi_jL_j+\frac{\bar{\theta}-(\delta^*+\pi_{k})}{\bar{\theta}-\underline{\theta}}q_{k}M\pi_{k}L_{k}.
\end{align}
\end{small}
According to Theorem 1, we can obtain the common equilibrium parameter
\begin{small}
\begin{align}\nonumber
\delta^*=\frac{[(M-1)(q_j+q_{k})+1](\bar{\theta}-\underline{\theta})-(M-1)[q_j(\pi_j-\underline{\theta})+q_{k}(\pi_{k}-\underline{\theta})]}{f^c (\bar{\theta}-\underline{\theta})+(M-1)(q_j+q_{k})}.
\end{align}
\end{small}
Then, for the optimal price of program $j$, the derivative of $U_B$ with respect to $\pi_j$ can be expressed as
\begin{small}
\begin{align}\nonumber
\frac{\partial U_B}{\partial\pi_j}=\frac{-(\frac{\partial \delta^*}{\partial\pi_j}+1)}{\bar{\theta}-\underline{\theta}}q_jM\pi_jL_j+\frac{\bar{\theta}-(\delta^*+\pi_j)}{\bar{\theta}-\underline{\theta}}q_jML_j+\frac{-\frac{\partial\delta^*}{\partial \pi_j}}{\bar{\theta}-\underline{\theta}}q_{k}M\pi_{k}L_{k},
\end{align}
\end{small}
where $\frac{\partial \delta^*}{\partial\pi_j}=\frac{-(M-1)q_j}{f^c (\bar{\theta}-\underline{\theta})+(M-1)(q_j+q_{k})}$.
By equating $\frac{\partial U_B}{\partial\pi_j}=0$, we have
\begin{small}
\begin{align}\label{opt1}
\pi_j^*=\frac{q_{k}(M-1)(L_j+L_{k})}{2L_j[q_{k}(M-1)+f^c(\bar{\theta}-\underline{\theta})]}\pi_{k}+\frac{A}{2[q_{k}(M-1)+f^c(\bar{\theta}-\underline{\theta})]},
\end{align}
\end{small}
where $A=f^c\bar{\theta}(\bar{\theta}-\underline{\theta})+(M-1)(q_j+q_{k})(\bar{\theta}-\underline{\theta})-[(M-1)(q_j+q_{k})+1](\bar{\theta}-\underline{\theta})$.

Similarly, we can obtain the optimal price for the cached program $k$, i.e.,
\begin{small}
\begin{align}\label{opt2}
\pi_{k}^*=\frac{q_{j}(M-1)(L_j+L_{k})}{2L_{k}[q_{j}(M-1)+f^c(\bar{\theta}-\underline{\theta})]}\pi_{j}+\frac{A}{2[q_{j}(M-1)+f^c(\bar{\theta}-\underline{\theta})]}.
\end{align}
\end{small}
By combining \eqref{opt1} and \eqref{opt2}, we have
\begin{small}
\begin{align}\label{finalopt1}
\pi_j^*=\frac{L_{k}[2L_jq_j(M-1)+2L_jf^c(\bar{\theta}-\underline{\theta})+q_{k}(M-1)(L_j+L_{k})]A}{4L_jL_{k}[q_jq_{k}(M-1)^2+f^c(\bar{\theta}-\underline{\theta})(M-1)(q_j+q_{k})+(f^c)^2(\bar{\theta}-\underline{\theta})^2]-q_jq_{k}(M-1)^2(L_j+L_{k})^2},
\end{align}
\end{small}
and
\begin{small}
\begin{align}\label{finalopt2}
\pi_{k}^*=\frac{L_{j}[2L_{k}q_{k}(M-1)+2L_{k}f^c(\bar{\theta}-\underline{\theta})+q_{j}(M-1)(L_j+L_{k})]A}{4L_jL_{k}[q_jq_{k}(M-1)^2+f^c(\bar{\theta}-\underline{\theta})(M-1)(q_j+q_{k})+(f^c)^2(\bar{\theta}-\underline{\theta})^2]-q_jq_{k}(M-1)^2(L_j+L_{k})^2}.
\end{align}
\end{small}
According to \eqref{finalopt1} and \eqref{finalopt2}, when $L_j=L_{k}$, we have $\pi_{j}^*=\pi_{k}^*=\frac{\bar{\theta}}{2}-\frac{1}{2f^c}$.
Besides, we calculate
\begin{small}
\begin{align}\nonumber
\pi_{j}^*-\pi_{k}^*=\frac{A[q_jL_j(L_{k}-L_j)+q_{k}L_{k}(L_{k}-L_j)]}{4L_jL_{k}[q_jq_{k}(M-1)^2+f^c(\bar{\theta}-\underline{\theta})(M-1)(q_j+q_{k})+(f^c)^2(\bar{\theta}-\underline{\theta})^2]-q_jq_{k}(M-1)^2(L_j+L_{k})^2}.
\end{align}
\end{small}
Therefore, when $L_j>L_{k}$, we have $\pi_{j}^*<\pi_{k}^*$. If $L_j<L_{k}$, $\pi_{j}^*>\pi_{k}^*$.

\section{Proof of Proposition 6.4} \label{appendicesF}
According to \eqref{finalopt1}, we calculate the first derivative with respect to $q_j$, i.e.,
\begin{small}
\begin{align}\nonumber
\frac{\partial\pi_j^*}{\partial q_j}=\frac{A(M-1)^2L_{k}(L_j-L_{k})(L_j+L_{k})q_{k}[2L_jf^c(\bar{\theta}-\underline{\theta})+(M-1)(L_j-L_{k})q_{k}]}{\bigg[4L_jL_{k}[q_jq_{k}(M-1)^2+f^c(\bar{\theta}-\underline{\theta})(M-1)(q_j+q_{k})+(f^c)^2(\bar{\theta}-\underline{\theta})^2]-q_jq_{k}(M-1)^2(L_j+L_{k})^2\bigg]^2}.
\end{align}
\end{small}
If $L_{j}>L_{k}$, we have $\frac{\partial\pi_j^*}{\partial q_j}>0, \forall q_{k}>0$. If $L_{j}<L_{k}$, when $0<q_{k}<\frac{2f^c(\bar{\theta}-\underline{\theta})L_j}{(L_{k}-L_{j})(M-1)}$, $\frac{\partial\pi_j^*}{\partial q_j}<0$. Otherwise, we have $\frac{\partial\pi_j^*}{\partial q_j}>0$.

\begin{footnotesize}
\bibliographystyle{IEEEtran}

\end{footnotesize}

\end{document}